\documentclass[sigconf]{acmart}

\usepackage{algorithm}
\usepackage{algorithmic}
\usepackage{subfigure}

\usepackage{threeparttable}

\usepackage{amsmath,amsfonts,bm}

\def\eqref#1{equation~\ref{#1}}

\def\1{\bm{1}}

\def\mL{{\bm{L}}}

\def\mP{{\bm{P}}}

\def\mR{{\bm{R}}}

\DeclareMathAlphabet{\mathsfit}{\encodingdefault}{\sfdefault}{m}{sl}
\SetMathAlphabet{\mathsfit}{bold}{\encodingdefault}{\sfdefault}{bx}{n}

\DeclareMathOperator*{\argmin}{arg\,min}

\usepackage[capitalise]{cleveref}

\newcommand{\archname}[0]{N\textsuperscript{3}H-Core} 
\newcommand{\archnames}[0]{N\textsuperscript{3}H-Core } 
\usepackage{tikz}
\newcommand{\ballnumber}[1]{\tikz[baseline=(myanchor.base)] \node[circle,fill=.,inner sep=1pt] (myanchor) {\color{-.}\bfseries\footnotesize #1};}
\begin{document}
\fancyhead{}
\title{N\textsuperscript{3}H-Core: \underline{N}euron-designed \underline{N}eural \underline{N}etwork Accelerator via FPGA-based \underline{H}eterogeneous Computing Cores}

\author{Yu Gong}
\authornote{Both authors contributed equally to this research. 
}
\orcid{1234-5678-9012}
\author{Zhihan Xu}
\authornotemark[1]
\affiliation{%
  \institution{Shanghai Qi Zhi Institute}
  \city{Shanghai}
  \country{China}
}
\email{{gongyu,xuzh}@sqz.ac.cn}

\author{Zhezhi He}
\authornote{Corresponding authors: Zhezhi He, and Li Jiang}
\affiliation{%
 \institution{Shanghai Jiao Tong University}
  \city{Shanghai}
  \country{China}
}
\email{zhezhi.he@sjtu.edu.cn}

\author{Weifeng Zhang}
\affiliation{%
  \institution{Alibaba Group US Inc.}
  \city{San Diego}
  \country{USA}  
  }
\email{weifeng.z@alibaba-inc.com}

\author{Xiaobing Tu}
\affiliation{%
  \institution{Alibaba Group}
  \country{China}
}
\email{xiaobing.tuxiaobin@alibaba-inc.com}

\author{Xiaoyao Liang}
\affiliation{%
  \institution{Shanghai Jiao Tong University}
  \city{Shanghai}
  \country{China}
}
\email{liang-xy@cs.sjtu.edu.cn}

\author{Li Jiang}
\authornotemark[2]
\affiliation{%
  \institution{Shanghai Jiao Tong University}
  \city{Shanghai}
  \country{China}
}
\email{ljiang_cs@sjtu.edu.cn}


\begin{abstract}
  Accelerating the neural network inference by FPGA has emerged as a popular option, since the reconfigurability and high performance computing capability of FPGA intrinsically satisfies the computation demand of the fast-evolving neural algorithms. 
  However, the popular neural accelerators on FPGA~(e.g., Xilinx DPU) mainly utilize the DSP resources for constructing their processing units, while the rich LUT resources are not well exploited. Via the software-hardware co-design approach, in this work, we develop an FPGA-based heterogeneous computing system for neural network acceleration. 
  From the hardware perspective, the proposed accelerator consists of DSP- and LUT-based GEneral Matrix-Multiplication~(GEMM) computing cores, which forms the entire computing system in a heterogeneous fashion. 
  The DSP- and LUT-based GEMM cores are computed w.r.t a unified Instruction Set Architecture~(ISA) and unified buffers.
  Along the data flow of the neural network inference path, the computation of the convolution/fully-connected layer is split into two portions, handled by the DSP- and LUT-based GEMM cores asynchronously.
  From the software perspective, we mathematically and systematically model the latency and resource utilization of the proposed heterogeneous accelerator, regarding varying system design configurations. 
  Through leveraging the reinforcement learning technique, we construct a framework to achieve end-to-end selection and optimization of the design specification of target heterogeneous accelerator, including workload split strategy, mixed-precision quantization scheme, and resource allocation of DSP- and LUT-core. 
  In virtue of the proposed design framework and heterogeneous computing system, our design outperforms the state-of-the-art Mix\&Match design with latency reduced by 1.12-1.32$\times$ with higher inference accuracy. The \archnames is open-sourced at: \url{https://github.com/elliothe/N3H_Core}.
\end{abstract}




\settopmatter{printacmref=false}
\setcopyright{none}
\renewcommand\footnotetextcopyrightpermission[1]{}
\pagestyle{plain}
\maketitle

\section{Introduction}
In the past decade, Deep Neural Networks~(DNN) has succeeded in various computer vision and natural language processing tasks~\cite{goodfellow2016deep}. 
Accelerating the DNN with domain-specific accelerator~\cite{chen2016eyeriss,chen2019eyeriss,guo2017angel} has been widely explored to counter the excellent computing power demanded by the mammoth DNN model that is growing exponentially~\cite{brown2020language,fedus2021switch}. 
Compared to other off-the-shelf hardware accelerators, FPGA can quickly adapt to the fast-evolving algorithms with outstanding performance~(e.g., latency, throughput, etc.), owing to its reconfigurability.  
It is known that the dominant operation in the DNN inference is the Multiplication and Accumulation ~(MAC)~\cite{chen2016eyeriss}. 
As the Digital Signal Processing~(DSP) units in the FPGA can perform MAC efficiently, prior FPGA-based DNN accelerator designs heavily use the DSP resources to construct the processing elements~(aka. \textit{DSP-core}) as DNN inference engine. 
In contrast, as another significant on-chip resource of FPGA, Look-Up Table (LUT) is merely utilized to build the peripheral computing units that perform the computations of batch-normalization, activation, etc.   

Thanks to recent advances in DNN compression algorithms~\cite{song2020drq,choi2018pact,esser2019learned,he2019simultaneously}, parameters of DNN can be converted from 32-bit floating point to extremely low bit-width~(e.g., $<$~4-bits) with negligible inference accuracy degradation, but significantly simplify the computation complexity and mitigate the on-/off-chip data access bottleneck~(aka. "memory wall")~\cite{wulf1995hitting,villa2014scaling}. 
Moreover, researchers are aware that the layers of DNN own different sensitivity to the quantization noise resulted from different bit-width. 
Thus, the DNN quantization with varying bit-width emerges as an alleged compression scheme to minimize the overall model size.
Note that, DSP-Core cannot efficiently support MAC operations with varying bit-width, due to the overhead caused by flexibility.
Although a recent design proposed by Yaman \textit{et. al.} called BISMO~\cite{umuroglu2018bismo} can efficiently support mixed-precision MACs via a bit-serial computing method, the inference latency of BISMO is still much higher than the pure DSP-based bit-parallel counterpart, to achieve the identical accuracy for low bit-width quantization.
For example, our preliminary experiment of 8-bit quantized ResNet-18 on ImageNet dataset~\cite{he2016deep,deng2009imagenet} shows the inference latency of DSP-based and LUT-based~(i.e., BISMO-based) design are 182ms and 309ms respectively. 
Thus, with the available DSP/LUT resources in the pool, an FPGA system that supports fully-flexible bit-width and outperforms other DSP-centric designs~\cite{guo2017angel} is desired. 

Furthermore, to maximize the performance of FPGA-based DNN accelerator, prior works~\cite{Zhang,Venieris2017} have conducted the design space exploration.
These studies either use the roof-line model to limit the design space or build the latency/performance model w.r.t the workload and architecture design parameters to explore the design space. 
While the former one cannot reflect the design parameters of the architecture, the latter does not impose constraints on parameters sufficiently, resulting in a enormous design space.

In summary, the preliminary architecture design of the FPGA-based DNN accelerator mainly counters the following drawbacks:
1) Rich on-chip LUT resource are not well utilized; 2) A high-performance architecture supporting mixed-precision operation is absent; 3) Lacking the systematic design methodology for the single-chip heterogeneous system.
As the countermeasure, in this work, we propose a heterogeneous computing architecture, which fully utilizes the on-chip resources~(including LUT, DSP, BRAM, etc.) for computing instead of control. 
Our proposed architecture consists of LUT-core and DSP-core that work jointly as a high-performance computing system. Such two computing cores handle computation with operands in flexible and fixed bit-width, respectively.
To ease the manpower to optimize the complicated system configuration on large design space, we \textbf{develop the end-to-end optimization framework as a systematic design methodology for the proposed single-chip heterogeneous system.}
Our contributions can be summarized as:
\begin{itemize}
    \item To fully exploit the on-chip resource of the target FPGA, we propose a heterogeneous DNN accelerator architecture called \archnames that consists of DSP- and LUT-centric computing units~(aka. DSP-core and LUT-core, respectively).
    
    \item For maximum throughput and minimum latency, We design the DSP- and LUT-core to operate w.r.t the predefined instructions in the intra-layer asynchronous fashion.
    
    \item We construct scalable and adaptable cost model across different DNNs and FPGA, to precisely estimate the resource utilization, inference latency, and other metrics-of-interests. 
    
    \item Through the cost model of the \archname, we apply the Reinforcement Learning~(RL) technique to build the end-to-end optimization framework that automatically generate the architecture configurations~(resource allocation), data flow~(layer-wise workload split ratio), and DNN~(quantization bit-width of weights and activations) respectively.
    Thus, given the target DNN and FPGA platform, an agile and optimal design can be rapidly achieved.
    
\end{itemize}

\section{Preliminary and Related Works}
\subsection{DNN Inference Acceleration on FPGA}
Recently, many studies~\cite{venieris2017latency,suda2016throughput,guo2017angel} have been conducted to explore using FPGA in DNN acceleration for flexibility and high computation performance. 
Those designs can be generally categorized into bit-parallel and bit-serial counterparts, specified as follow.

\subsubsection{Bit-parallel Acceleration} 
For bit-parallel computing, the bits of operands are fed in parallel into the computing unit, e.g., the DSP on FPGA that typically used to compute multiplications.
Prior designs~\cite{venieris2017latency,suda2016throughput,guo2017angel,9407035} focus on bit-parallel accelerator with DSP for optimized latency, throughput, etc.
For latency reduction, \cite{venieris2017latency} proposes a latency-centric optimizer to explore design space efficiently. 
\cite{suda2016throughput} presents a systematic design space exploration methodology to maximize the throughput of an OpenCL-based FPGA accelerator, taking source available on-chip as constraints.
\cite{9407035} proposes a heterogeneous architecture that utilizes DSP and LUT resources.
The logarithmic quantization is used to replace the multiplication with bit-shift.
However, the architecture of~\cite{9407035} is not bit-width flexible and without automated architecture optimization.

\begin{figure}[t]
    \centering
    \includegraphics[width=\columnwidth]{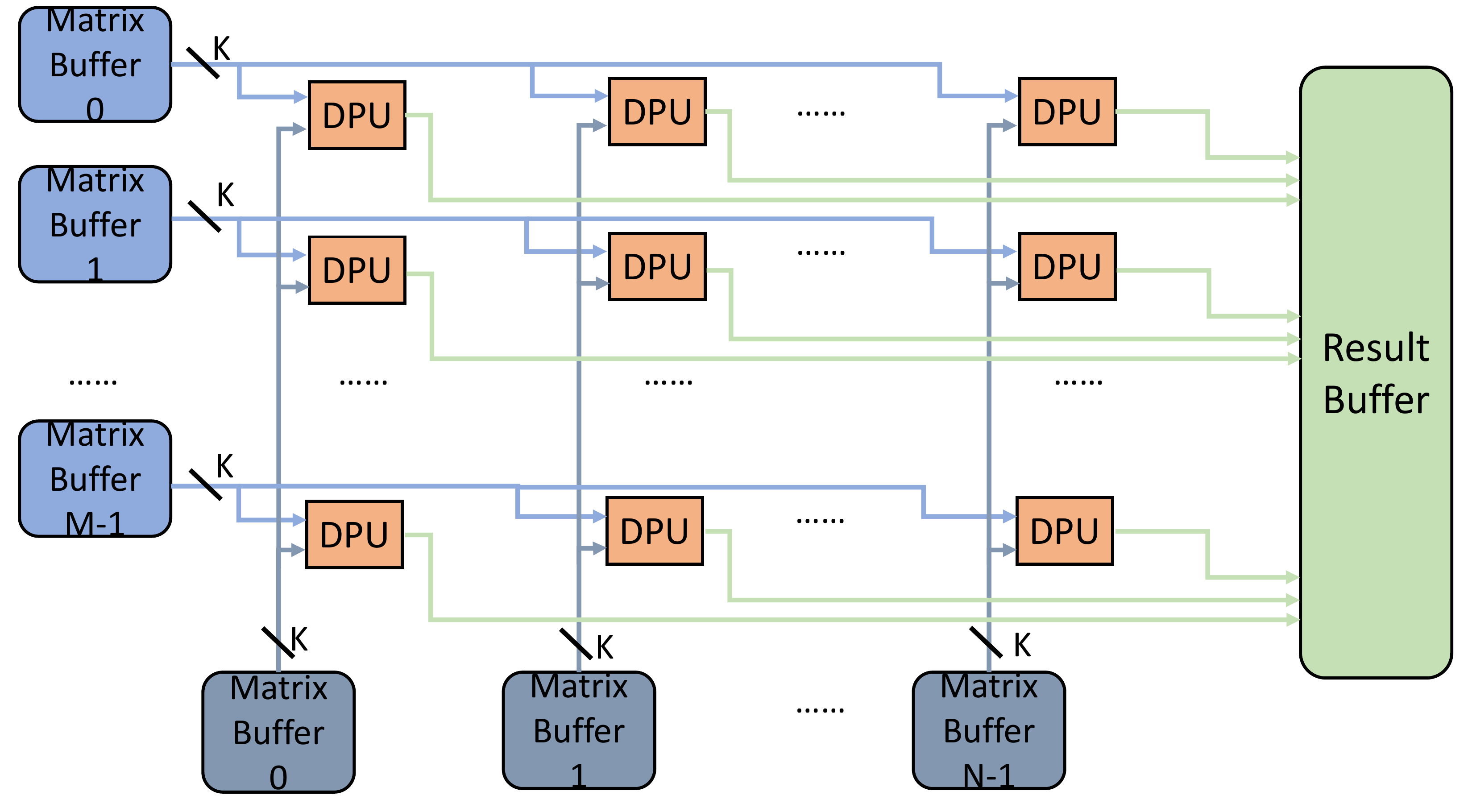}
    \caption{Overview of bit-serial GEMM core in BISMO~\cite{umuroglu2018bismo}. }
    \label{fig:BISMO architecture}
\end{figure}

\subsubsection{Bit-serial Acceleration}
\label{sec:BISMO introduction}

In bit-parallel scheme, accelerator is designed to support highest operand bit-width, even when some operands are in low bit-width, thus lead to great hardware overhead.
To mitigate that, bit-serial scheme computes one binary multiplication per cycle which can compose operations on higher bit-width operands. 
Given integer matrix $\mL$ and $\mR$, their multiplication is:
\begin{small}
\begin{equation}
\label{equ: bit-serial multiplication}
\begin{aligned}
\mL &=\left[\begin{array}{ll}
2 & 0 \\
1 & 3
\end{array}\right]=2^{1} \mL^{[1]}+2^{0} \mL^{[0]}=2^{1}\left[\begin{array}{ll}
1 & 0 \\
0 & 1
\end{array}\right]+2^{0}\left[\begin{array}{ll}
0 & 0 \\
1 & 1
\end{array}\right] \\
\mR &=\left[\begin{array}{ll}
0 & 1 \\
1 & 2
\end{array}\right]=2^{1} \mR^{[1]}+2^{0} \mR^{[0]}=2^{1}\left[\begin{array}{ll}
0 & 0 \\
0 & 1
\end{array}\right]+2^{0}\left[\begin{array}{ll}
0 & 1 \\
1 & 0
\end{array}\right] \\
\mP &= \mL \cdot \mR=\left(2^{1} \mL^{[1]}+2^{0} \mL^{[0]}\right) \cdot\left(2^{1} \mR^{[1]}+2^{0} \mR^{[0]}\right) \\
&=2^{2} \mL^{[1]} \cdot \mR^{[1]}+2^{1} \mL^{[1]} \cdot \mR^{[0]}+2^{1} \mL^{[0]} \cdot \mR^{[1]}+2^{0} \mL^{[0]} \cdot \mR^{[0]}
\end{aligned}
\end{equation}
\end{small}
where $\mL\times \mR$ is decomposed into weighted sum of the binary matrix multiplication. Such computation transform is hardware-friendly and can be efficiently implemented with LUT on FPGA, which is adopted in BISMO design~\cite{umuroglu2018bismo}.
Based on~\cref{equ: bit-serial multiplication}, BISMO propose a GEMM core~(depicted in Figure~\ref{fig:BISMO architecture} composed of a M-by-N DPU array, where each DPU unit performs the binary vector dot-product via the \texttt{XNOR} and \texttt{pop-count}.
According to the bit-serial computing fashion, the computing latency of BISMO is proportional to the operand bit-width of $\mL$ and $\mR$. Thus, such computing architecture is prone to compute with the low bit-width operands.

\subsection{Model Quantization}
DNN model quantization has emerged as a mandatory technique for high-performance DNN inference.
Thanks to the advances in model quantization algorithm~\cite{zhou2016dorefa, choi2018pact, choukroun2019low,he2019optimize,he2019simultaneously,liu2021improving}, the activations and weights in 32-bit floating-point ({\tt fp32}) can be quantized into extreme low bit-width with negligible inference accuracy degradation, using uniform~\cite{jacob2018quantization} or non-uniform quantizer~\cite{miyashita2016convolutional} in quantization-aware training~\cite{he2019optimize,he2019simultaneously} or post-training quantization~\cite{liu2021improving}.
In this work, we focus on the model quantization using $N_\textrm{bits}$-bit uniform quantizer, and its quantization function can be expressed as:
\begin{equation}
\label{quant}
\hat{x} = f_\textrm{q}(x,s)= \textrm{clip} \big{(} \textrm{round}(\dfrac{x}{s}),\hat{\alpha},\hat{\beta} \big{)}    
\end{equation}
where $s$ is the step size. $\hat{\alpha} = -2^{N_\textrm{bits}-1}$ and $\hat{\beta}=2^{N_\textrm{bits}-1}-1$ are the upper and lower bound for the $N_{\textrm{bits}}$ integer representation. $\textrm{clip}(\cdot,\hat{\alpha},\hat{\beta}) = \min( \max(\cdot, \hat{\beta}),\hat{\alpha})$ is the clipping function, and $\textrm{round}()$ denotes the round function. 
With the uniform quantizer adopted, two quantization schemes~(uniform precision and mixed precision) are usually applied upon the entire DNN mode.

\textbf{Uniform Precision.}
Uniform precision quantization assigns identical bit-width to each layer of target DNN. 
Researchers~\cite{zhou2016dorefa,choukroun2019low,gong2019differentiable} have shown that {\tt INT8}, even {\tt INT4}, can retain accuracy on many neural networks. 
For the general-purpose computing hardware~(e.g., CPU or GPU), due to the hardware constraint, only {\tt INT4} and {\tt INT8} are supported.
Thus, uniform precision quantization is a relatively general method, as most hardware platforms only support computation~(e.g., MAC) in specific bit-width. 

\textbf{Mixed Precision.} As the prior investigations~\cite{cai2020zeroq,wu2018mixed,wang2019haq} reveal the fact that layers of DNN own different redundancy, the quantization with varying bit-width on different layers emerges as a potential scheme to minimize the model size of target DNN.  
\cite{zhu2018adaptive} determines the layer-wise bit-width based on the entropy of weights and activations. 
\cite{dong2019hawq} utilizes the second-order information, i.e., Hessian matrix, to determine the bit-width, where the block with higher top Hessian eigenvalue will be assigned with higher bit-width. 
Elthakeb \textit{et al.}~\cite{elthakeb2019releq} use the sample efficiency of Proximal Policy Optimization~(PPO) to find the optimal bit-width, which is the first to use RL for quantization. 
Wang \textit{et al.}~\cite{wang2019haq} predict the layer-wise bit-width via Deep Deterministic Policy Gradient~(DDPG) algorithm with the target hardware taken into consideration.

\section{Architecture of \archname}
\label{sec:arch}

In this work, we propose a heterogeneous neural network accelerator called \archname, which fully exploits both the DSP and LUT resources in a heterogeneous fashion. Its architecture, data flow and cost and latency models are specified as follows.


\subsection{Architecture Overview of \archname}
\label{subsec:arch_overview}

\begin{figure}[t]
    \centering
    \includegraphics[width = .97\columnwidth]{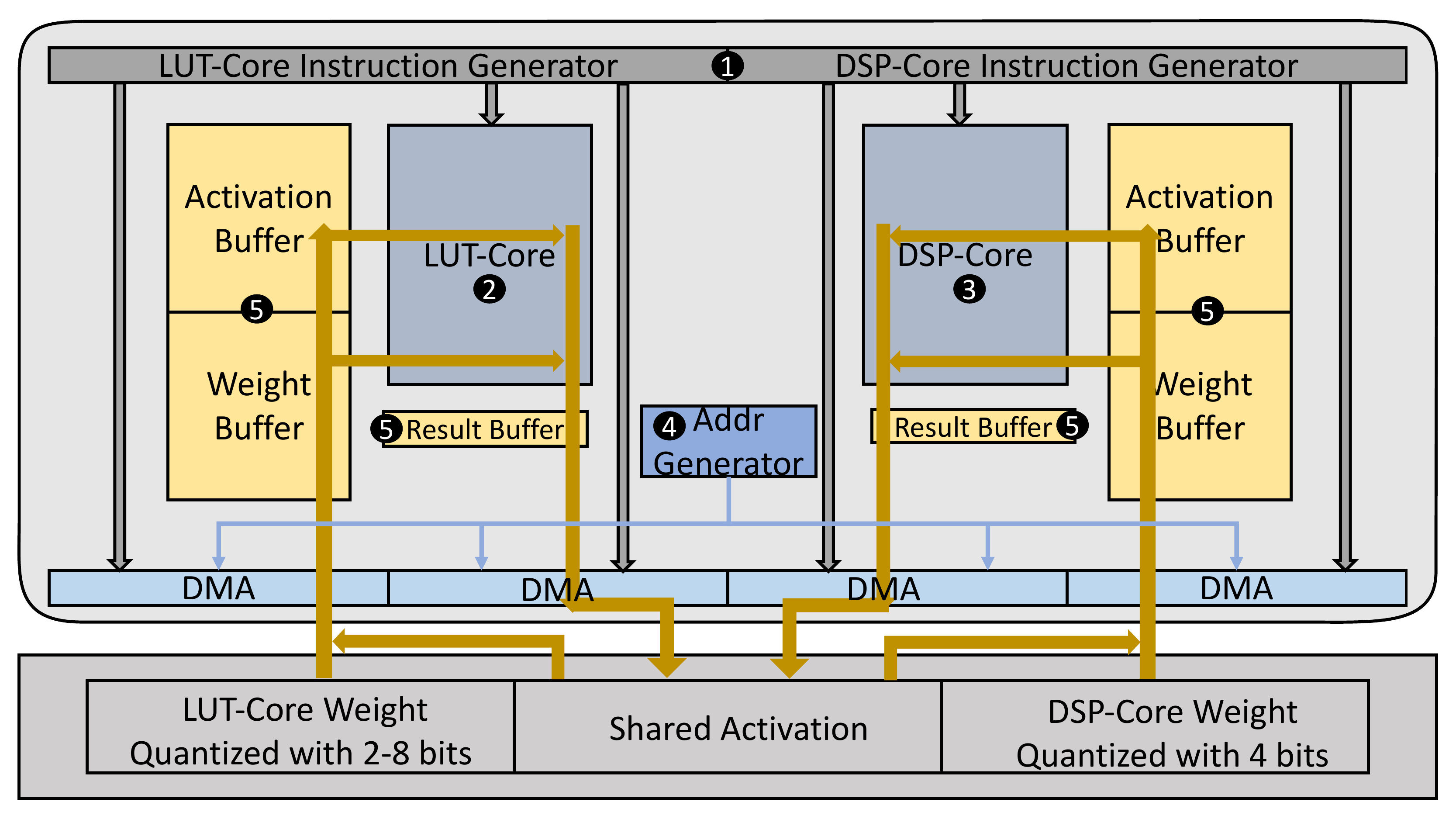}
    \caption{Overview of proposed \archname architecture, which includes the DSP- and LUT-core. The grey and green lines are the instruction flow and data flow respectively.}
    \label{fig:arch_overview}
\end{figure}

As depicted in~\cref{fig:arch_overview}, \archnames architecture consists of \ballnumber{1}~instruction generator, heterogeneous computing cores (i.e., \ballnumber{2}~LUT-core and \ballnumber{3}~DSP-core), \ballnumber{4}~address generator, \ballnumber{5}~peripheral buffers.

\ballnumber{1}~\textbf{Instruction Generator.} 
\archnames is an intra-layer asynchronous and inter-layer synchronous instruction-based neural network accelerator.
Thus, the instructions adopted can be generally separated into two categories: operation instruction and synchronization instruction, which jointly work for accurate pipelining computation within LUT- and DSP- core. 
Within each layer, \textit{Operation Instructions} are generated asynchronously and specifically for the LUT-Core and DSP-Core, consisting of three types \{{\tt Fetch}, {\tt Execute}, {\tt Result}\} that corresponds to three operating phases of LUT- and DSP-core, respectively.
1) In {\tt Fetch} phase, Direct-Memory Access~(DMA) module loads weight and activation data from the off-chip DDR to on-chip buffers;
2) In {\tt Execute} phase, data is transmitted from buffers to computing cores for multiplication and accumulation;
3) In {\tt Result} phase, the calculation results from the LUT- and DSP-core are written back to the shared result buffer.
Thanks to the synchronous instructions~(\texttt{Sync}), two cores are synchronized after they complete the assigned workloads, then launch the computation of successive layer without conflicts.

\begin{figure}[t]
    \centering
    \subfigure[Instruction synchronization via queue.]{
    \includegraphics[width=\columnwidth]{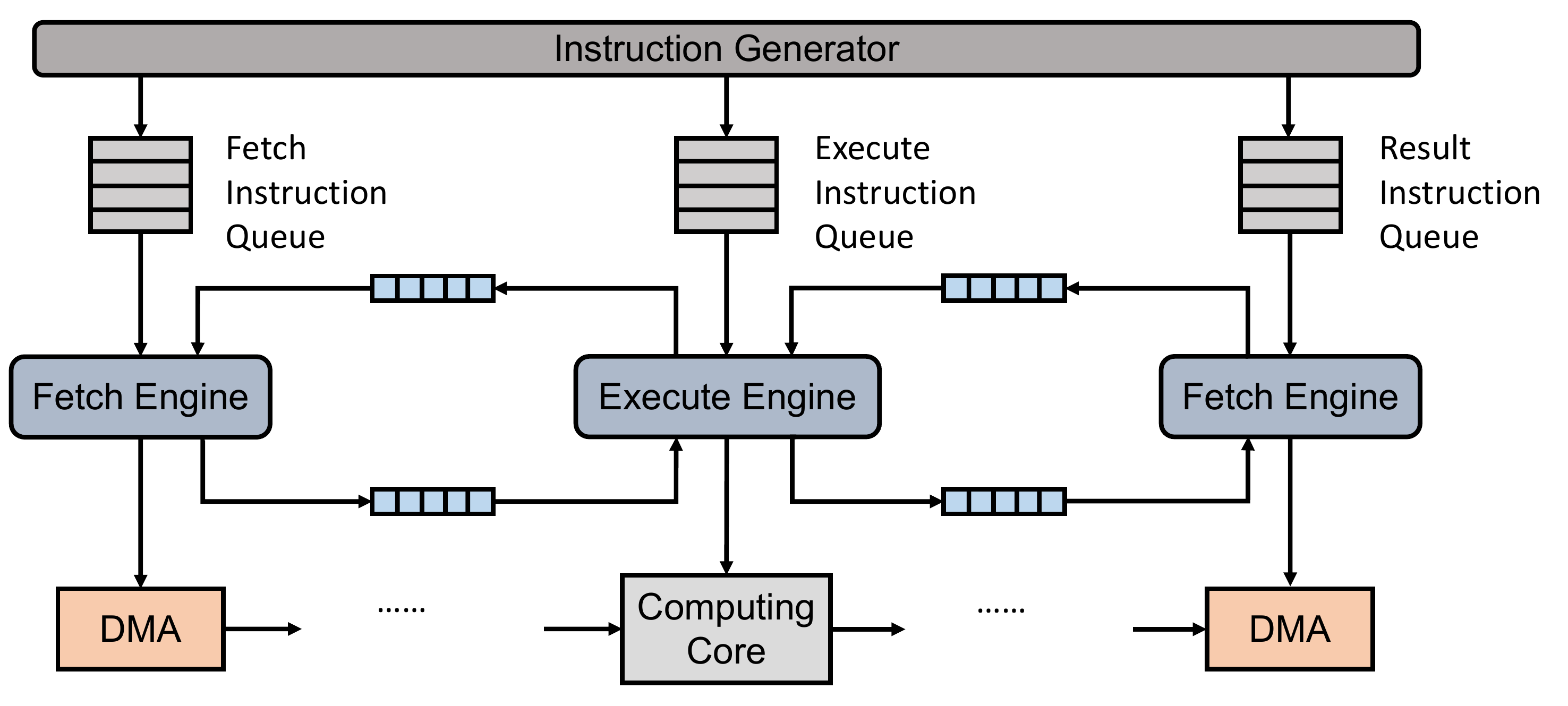}
    }
    \subfigure[Timeline of scheduled instructions.]{
    \includegraphics[width=\columnwidth]{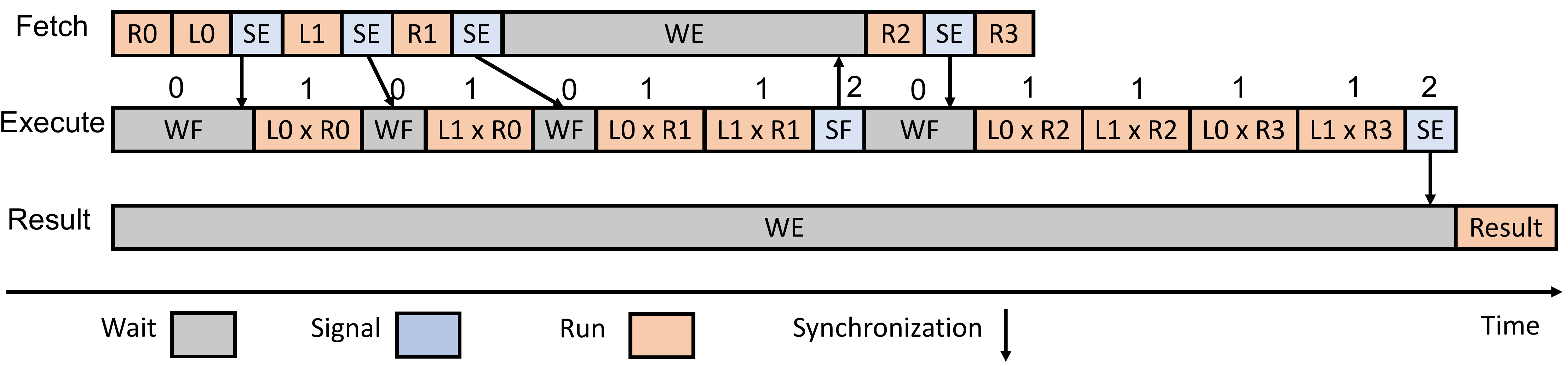}
    } 
    \caption{Overview of instruction synchronization and scheduling in \archname.}
    \label{fig: Instructionn schedule}
\end{figure}

Based on prior designs~\cite{umuroglu2018bismo,Matam2013}, we propose a new schedule for matrix multiplication and design \texttt{Fetch}, \texttt{Execute} and \texttt{Result} engines to run the corresponding instructions, as described in~\cref{fig: Instructionn schedule}.
Referring to~\cref{equ: bit-serial multiplication}, we assume the activation buffers possess the capacity of the activation matrix $\mL$, and weight buffer is the half capacity of weight matrix $\mR$. 
To begin with the first part\footnote{For simplicity, we write decomposed $\mR^{[0]}$ and $\mL^{[0]}$ in~\cref{equ: bit-serial multiplication} as R0 and L0 respectively} of $\mR$ and $\mL$, R0 is fetched by fetch engine, then $L0$ is fetched as well. 
Meanwhile, execute and fetch engines are in waiting status, i.e., Wait-Fetch~(WF) and Wait-Execute~(WE).
When fetch instructions is completed, the fetch engine sends a synchronization signal to activate the execute engine, i.e., Sync-Execute~(SE).
After the execute engine receives the signal, it immediately transfer from WF status to run the execute instruction of L0$\times$R0.
While the L0$\times$R0 is executing, the fetch engine start to load following binary matrix L1 for pipeling.
After fetching the first tile of weight data R0 and R1, the weight buffer is filled, so the fetch engine is in WE status, waiting for a synchronization signal from execute engine.
After the signal is received, the fetch engine continues to fetch the next tile of weight data R2 and R3.
Once the R2 is fetched, the execute engine multiplies the pre-loaded binary matrices of $\mL$ in a sequence. 
When all the multiplications are done, the result engine receives the synchronization signal~(Sync-Execute, SE) from the execute engine, then shift from Wait-Execute~(WE) status to run \texttt{Result} instruction that store the computation result back to DDR. 
In this way, the design bridges the gap between data movement and processing.

For instruction compatibility for DSP- and LUT-core, all instructions are 128-bits composing different segments for information representation. 
For \texttt{Fetch} and \texttt{Result} instructions, they consist of base address~(16-bits), stage control~(3-bits), read
/write range~(1-bit) of on-chip buffer and base address~(32-bits), offset~(24-bits), and write/read range~(16-bits) of the DDR. 
For \texttt{Execute} instruction, it includes the on-chip buffer data address, and the GEMM-Core parameters listed in~\cref{table:params_to_optimize}.
For \texttt{sync} instruction, it includes the current~(1-bit) and next state~(2-bits) of each Engine and the flag~(3-bits) indicating the sent of synchronization token.

\textbf{Heterogeneous Computing Core.} 
To fully utilize the on-chip resource of a target FPGA for boosted computing power, we develop a heterogeneous computing core that consists of \ballnumber{2}~LUT-core and \ballnumber{3}~DSP-core. 
Given a computing task~(e.g., inference with one layer of DNN), the computation workload is split w.r.t a designated ratio, then performed by the LUT- and DSP-core asynchronously. 
Note that, the optimization of workload split ratio will be discussed in~\cref{5.3}.
For the LUT-core, we adopt the open-sourced BISMO\footnote{BISMO: \url{https://github.com/EECS-NTNU/bismo}}~\cite{umuroglu2018bismo} as its backbone, which is a GEMM kernel with run-time adjustable operand bit-width in virtue of the bit-serial computing mechanism. 
In contrast to the LUT-core operating in the bit-serial fashion, the DSP-core is developed for bit-parallel computing. 
Overall, the heterogeneous computing core incorporates prior-neglected rich LUT resources and introduces a more flexible weight-activation quantization scheme for accuracy lossless mapping.    

\ballnumber{5}~\textbf{Peripheral Buffer.} 
To avoid the frequent off-chip memory access that hampers the overall inference latency and improve the throughput via pipelining, the proposed architecture allocates each computing core with input/weight/result buffers. 
Weight and activation data are fetched via Direct Memory Access~(DMA) to the dedicated buffers, then transferred to the computing cores. 
Once the computation is done, the result stored in the result buffer is then written to DDR as the activation of the next layer.

\textbf{Memory Control.} 
Two workload partitions allocated to two computing cores are stored at different regions on DDR to make the data access more efficiently and avoid conflict.
\ballnumber{4}~Address Generator controls the base address and strides to instruct two partitions fetched from DDR.
After the computation of one layer is finished, \ballnumber{4} gives new base addresses and strides to store the result. 

\begin{table}[t]
\caption{Architecture design knobs for \archname.}
\label{table:params_to_optimize}
\begin{tabular}{@{}cl@{}}
\toprule
\textit{Notation} & \textit{Description}                           \\ \midrule
$M$                 & Numbers of rows in LUT-Core (DPU array)                    \\
$N$                 & Numbers of columns in LUT-Core (DPU array)                          \\
$K$                  & Input bit width of DPU               \\
$D_{\textrm{L,buf}}^{\textrm{a}}$                 & \textbf{D}epth of \textbf{a}ctivation \textbf{buf}fers in \textbf{L}UT-Core        \\
$D_{\textrm{L,buf}}^{\textrm{w}}$                 & \textbf{D}epth of \textbf{w}eight \textbf{buf}fers in \textbf{L}UT-Core  
\\ \midrule
$N_{\textrm{reg,row}}^{\textrm{a}}$
& \textbf{N}umbers of \textbf{row}s in \textbf{a}ctivation \textbf{reg}ister array   \\
$N_{\textrm{reg,col}}^{\textrm{a}}$
& \textbf{N}umbers of \textbf{col}umns in \textbf{a}ctivation \textbf{reg}ister array \\
$N_{\textrm{reg,col}}^{\textrm{w}}$
& \textbf{N}umbers of \textbf{col}umns in \textbf{w}eight \textbf{reg}ister array    \\
$D_\textrm{D,buf}^{\textrm{a}}$               & \textbf{D}epth of \textbf{a}ctivation \textbf{buf}fers in \textbf{D}SP-Core        \\
$D_\textrm{D,buf}^{\textrm{w}}$               & \textbf{D}epth of \textbf{w}eight \textbf{buf}fers in \textbf{D}SP-Core 
\\ \midrule
$B^{\textrm{a}}$                &  \textbf{B}it-widths of quantized activation
\\
$B^{\textrm{w-L}}$                & \textbf{B}it-widths of quantized \textbf{w}eight on \textbf{L}UT-Core
\\
$B^{\textrm{w-D}}$                & \textbf{B}it-widths of quantized \textbf{w}eight on \textbf{D}SP-Core
\\ \bottomrule
\end{tabular}
\end{table}

\subsection{Heterogeneous Computing Core}
\subsubsection{LUT-Core}
Through the conversion function~(i.e., {\tt im2col}), the computation of various parametric layers~(e.g., convolution layer, fully-connected layers) can all be computed via the matrix multiplication~(GEMM).
\cref{fig:BISMO architecture} illustrates the architecture of LUT-Core. 
In LUT-Core, the buffers allocated to the row are designated as the activation buffer~(indexed from $0$ to $M-1$ in~\cref{fig:BISMO architecture}), the activation buffers have the depth of $D_{\textrm{L,buf}}^{a}$, whose interface width is $K$. 
Similarly, the buffer allocated to the column is weight buffers indexed from $0$ to $N-1$, and the depth is $D_{\textrm{L,buf}}^{W}$, whose interface width is $K$ as well. 
Each DPU reads $K$-bits from the activation buffers and $K$-bits from weight buffers simultaneously, then store the binary dot multiplication results as partial sum.
The LUT-Core can process $M \times K \times N$ bits in parallel. 
As listed in~\cref{table:params_to_optimize}, $\{ M,N,K,D_{\textrm{L,buf}}^{a},D_{\textrm{L,buf}}^{w}\}$ are five important configuration parameters significantly affect the latency, which will be explored at design time. 

\subsubsection{DSP-Core}
\label{sec:DSP-Core}

\begin{figure}[t]
\centering      
\includegraphics[width=.95\columnwidth]{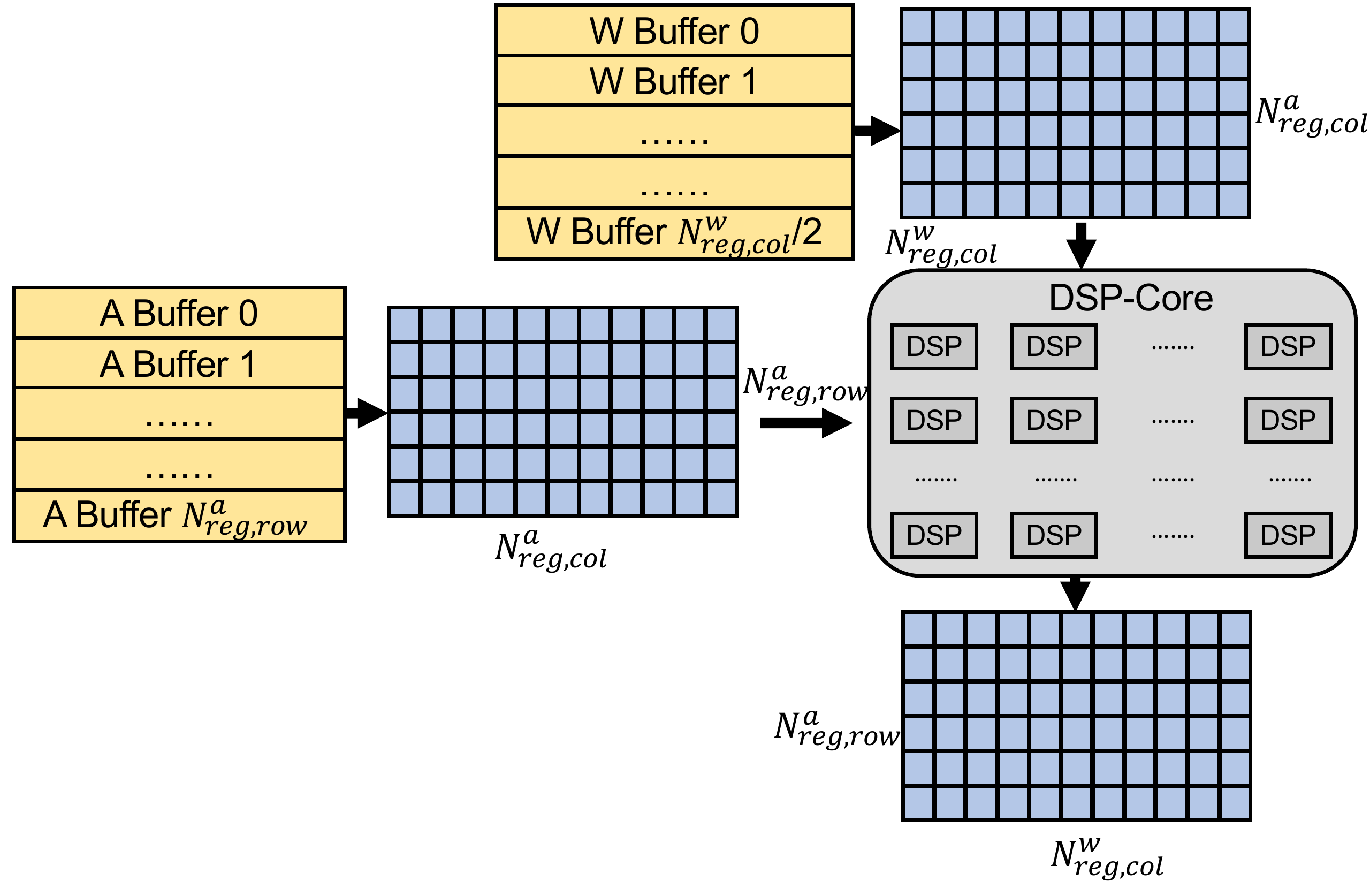}
\caption{DSP-Core architecture. Activation and weight data are computed in tiling fashion via DSP array.}
\label{fig:DSP Array}
\end{figure}

The architecture detail of DSP array is illustrated in~\cref{fig:DSP Array}. Note that, the DSP core performs the GEMM in tiling fashion as well.
In {\tt Execute} phase, a tile of input activation data and weight data are read from buffers into register arrays. 
The register array of activation data has $N_{\textrm{reg,row}}^{\textrm{a}}$ rows and $N_{\textrm{reg,col}}^{\textrm{a}}$ columns. 
Correspondingly, the register array of weight data has $N_{\textrm{reg,col}}^{\textrm{a}}$ rows\footnote{Note that, the row size of weight register array is identical to the column size of activation register array.} and $N_{\textrm{reg,col}}^{\textrm{w}}$ columns. 
For activation buffer, the size of each row of activation register array is same as the activation buffer.
Thus, the total bit-width of each buffer is $N_{\textrm{reg,col}}^{\textrm{a}} \times B^{\textrm{a}}$, where $B^{\textrm{a}}$ is the activation bit-width.
During the computing, each row of the activation array is filled with an activation buffer per cycle.
For weight buffer, the bit-width of each buffer is $N_{\textrm{reg,col}}^{\textrm{a}} \times B^{\textrm{w-D}}$, and each two columns of the weight register array is filled with a weight buffer, thus it requires two cycles to fill the weight register array with weight buffer.

\subsection{Cost and Latency Model}

To minimize the latency by finding the optimal architecture configuration, we build the cost and latency models here for rapid design space exploration.
The cost model simulates the resource utilization of LUT, BRAM, and DSP on FPGA, while the architecture configuration in~\cref{table:params_to_optimize} is constrained by available on-chip resource.
Besides, the latency model describes the end-to-end inference latency. 

\textbf{DSP-Core Cost Model.} 
As described in~\cref{sec:DSP-Core}, the number of activation buffer equals to the $N_{\textrm{reg,row}}^{\textrm{a}}$, and each buffer has the depth of $D_{\textrm{buf}}^{\textrm{a}}$ with the interface of $N_{\textrm{reg, col}}^{\textrm{a}} \times B^{\textrm{a}}$.
Since our DSP-core is designed for 4-bit multiplication, the quantized activation bit-width ranges from 2 to 4 bits. 
When the activation bit-width $B^{\textrm{a}}$ is less than 4, we pad zeros to make it 4-bits for storing in buffers.
The BRAMs forming buffers are 1024 in-depth and 36-bits wide, and we use the 32-bits for convenience. 
Thus, the BRAM size consumed by DSP-Core is formulated as:
\begin{equation}
\label{equ: DSP-Core cost model of BRAM}
\begin{split}
    \textrm{BRAM}_{\textrm{D-core}} (N_{\textrm{reg,row}}^{\textrm{a}},N_{\textrm{reg,col}}^{\textrm{a}},N_{\textrm{reg,col}}^{\textrm{w}},D_{\textrm{D,buf}}^{\textrm{a}},D_{\textrm{D,buf}}^{\textrm{w}})= 
    \\
    \lceil \dfrac{N_{\textrm{reg,row}}^{\textrm{a}} \cdot 4}{32} \rceil \cdot (N_{\textrm{reg,col}}^{\textrm{a}} \cdot \lceil \dfrac{ D_{\textrm{D,buf}}^{\textrm{a}} }{1024} \rceil + \dfrac{ N_{\textrm{reg,col}}^{\textrm{w}} }{2} \cdot \lceil \dfrac{ D_{\textrm{D,buf}}^{\textrm{w}} }{1024} \rceil)
\end{split}
\end{equation}
where $ \lceil  N_{\textrm{reg,col}}^{\textrm{a}} \times 4  / 32 \rceil \cdot  \lceil D_{\textrm{D,buf}}^{\textrm{a}}/1024 \rceil $ is the BRAM size that one activation buffer consumes, and $N_{\textrm{reg,row}}^{\textrm{a}}$ is the size of activation buffers. 
Likewise, since we allocate data in one buffer to two columns of the weight register array, the amount of weight buffers is $N_{\textrm{reg,col}}^{\textrm{w}}/2$, and $\lceil  N_{\textrm{reg,col}}^{\textrm{a}} \times 4 / 32 \rceil \cdot \lceil D_{\textrm{D,buf}}^{\textrm{w}} / 1024 \rceil $ is the BRAM size that one weight buffer consumed.
For LUT usage of DSP-Core, it is mainly consumed by the instruction generator and the control module, which is about 1000 as a constant~($\textrm{LUT}_{\textrm{D-core}}=1000$). 
The utilization rate of DSP resources is predefined as 100\% at design time.
The DSP-Core utilizes all the DSP resource available on chip, which means $\textrm{DSP}_{\textrm{D-core}}=\textrm{DSP}_{\textrm{Available}}$.

\textbf{LUT-Core Cost Model.}
Since we adopt the BISMO~\cite{umuroglu2018bismo} as the backbone of LUT-core, we inherits its cost model as:
\begin{equation}
\label{equ: LUT-Core cost model of LUT}  
    \textrm{LUT}_{\textrm{L-core}}(M,K,N)  = M \cdot N \cdot {\left( aK + b + c \right)} + d
\end{equation}
\begin{equation}
\label{equ: LUT-Core cost model of BRAM}
\begin{split}
    \textrm{BRAM}_{\textrm{L-core}} = \lceil \dfrac{K}{32} \rceil \cdot (M \cdot \lceil \dfrac{D_\textrm{L,buf}^{a}}{1024} \rceil + N \cdot \lceil\frac{D_\textrm{L,buf}^{w}}{1024}\rceil)
\end{split}
\end{equation}
where $\{a,b,c,d\}=\{1.17, 120.1, 44.1, 718\}$ are the fitting coefficient. 
From~\cref{equ: LUT-Core cost model of LUT}, the LUT utilization is proportional to $M$, $K$, and $N$.
We use 1024 in-depth and 36-bits wide BRAMs to build the buffer, and we only use 32-bits of the original interface. 
As mentioned before, there are $M$ activation buffers and each of them is comprised of $\lceil K/32 \rceil \cdot \lceil D_\textrm{L,buf}^{a}/1024 \rceil$ BRAMs. 
Similarly, the BRAMs consumed by weight buffer is calculated by~\cref{equ: LUT-Core cost model of BRAM}.

\textbf{DSP-Core Latency Model.} 
As the DSP-core is an instruction-oriented computing unit whose instruction pipeline is organized in a specific pattern, we build its latency model w.r.t its instruction pipeline.
We focus on the execute engine and count the latency as following queues: $L_{\textrm{wait}}$, $L_{\textrm{run}}$,$L_{\textrm{rst}}$ and $L_{\textrm{sig}}$. These four parts of latency is relevant to the parameters \{$N_{\textrm{reg,row}}^{\textrm{a}}$, $N_{\textrm{reg,col}}^{\textrm{a}}$, $N_{\textrm{reg,col}}^{\textrm{w}}$, $D_{\textrm{D,buf}}^{\textrm{a}}$, $D_{\textrm{D,buf}}^{\textrm{a}}$ \} in~\cref{table:params_to_optimize}.
Then, we trace the instructions streams and accumulate the latency of each phase on execute engine.
Note that, $L_{\textrm{wait}}$ is the latency of waiting for fetching data~(i.e., phases denoted by 0 in~\cref{fig: Instructionn schedule}), $L_{\textrm{run}}$ is the latency of execution~(phases denoted by 1), $L_{\textrm{sig}}$ is the latency of sending signal~(indicated by 2), and $L_{\textrm{rst}}$ is the latency of result phase. 
By accumulating each latency, the computational latency of one layer is formulated as:  
\begin{equation}
    \centering
    \label{equ: DSP-Core latency model}
    \begin{split}
        L_{\textrm{DSP}} = g( N_{\textrm{reg,row}}^{\textrm{a}}, N_{\textrm{reg,col}}^{\textrm{a}}, N_{\textrm{reg,col}}^{\textrm{w}}, D_{\textrm{D,buf}}^{\textrm{a}}, D_{\textrm{D,buf}}^{\textrm{a}})=
        \\
        \sum{L_{\textrm{wait}}}+\sum{L_{\textrm{run}}}+\sum{L_{\textrm{sig}}}+\sum{L_{\textrm{rst}}}    
    \end{split}
\end{equation}
To fully utilize the DSP resource on-chip, we adopt a simple simulation method that $N_{\textrm{reg,col}}^{\textrm{a}}$, $N_{\textrm{reg,col}}^{\textrm{w}}$ are fixed to 16, and only $N_{\textrm{reg,row}}^{\textrm{a}}$ can be modified that determines the latency performance. 
Therefore, it ensures that DSPs are fully-utilized by modifying only one parameter, and a simple core design method is provided at the same time. 
Then, the calculation of~\cref{equ: DSP-Core latency model} can be simplified as:
\begin{equation}
    \centering
    \label{equ: DSP-Core latency model simple}
    L_{\textrm{DSP}} = g( N_{\textrm{reg,row}}^{\textrm{a}}, D_{\textrm{D,buf}}^{\textrm{a}}, D_{\textrm{D,buf}}^{\textrm{a}})
\end{equation}

\textbf{LUT-Core Latency Model.} 
Likewise, the latency model of LUT-Core is built in the same manner via describing and simulating the instruction streams. 
However, the main difference is that LUT-Core computes the MACs in a bit-serial fashion, whose latency is determined by the bit-width of activation~($B^{\textrm{a}}$) and weight ($B^{\textrm{w-L}}$) matrices.
Thus, the parameters that relevant to the LUT-Core latency is \{$B_{\textrm{bits}}^{\textrm{a}}$, $B_{\textrm{L,bits}}^{\textrm{w}}$, $M$, $K$, $N$, $D_{\textrm{L,buf}}^{\textrm{a}}$, $D_{\textrm{L,buf}}^{\textrm{w}}$ \}, and the latency model of LUT-Core is formulated as:
\begin{equation}
    \centering
    \label{equ: LUT-Core latency model}
    \begin{split}
    L_{\textrm{LUT}} = f(B_{\textrm{bits}}^{\textrm{a}}, B_{\textrm{L,bits}}^{\textrm{w}}, M, K, N, D_{\textrm{L,buf}}^{\textrm{a}}, D_{\textrm{L,buf}}^{\textrm{w}}) =
    \\
    \sum{L_{\textrm{wait}}}+\sum{L_{\textrm{run}}}+\sum{L_{\textrm{sig}}}+\sum{L_{\textrm{rst}}}      
    \end{split}
\end{equation}
Note that, the weight buffer reads one tile at once.
Thus, the capacity of the weight buffer is the size of one tile weight data, while the depth change will not affect the latency.
Correspondingly, we simplify the latency model of LUT-Core as:
\begin{equation}
    \label{equ: LUT-Core latency model simple}
        L_{\textrm{LUT}} = f(B_{\textrm{bits}}^{\textrm{a}}, B_{\textrm{L,bits}}^{\textrm{w}}, M, K, N, D_{\textrm{L,buf}}^{\textrm{a}}) 
\end{equation}
where such simplification also shrink the design space to be explored by reducing the variation of $D_{\textrm{L,buf}}^{\textrm{w}}$.

\begin{figure}[t]
    \centering
    \subfigure[Predicted latency vs actual latency.]{
    \label{fig:a}
    \includegraphics[width = 0.46\columnwidth]{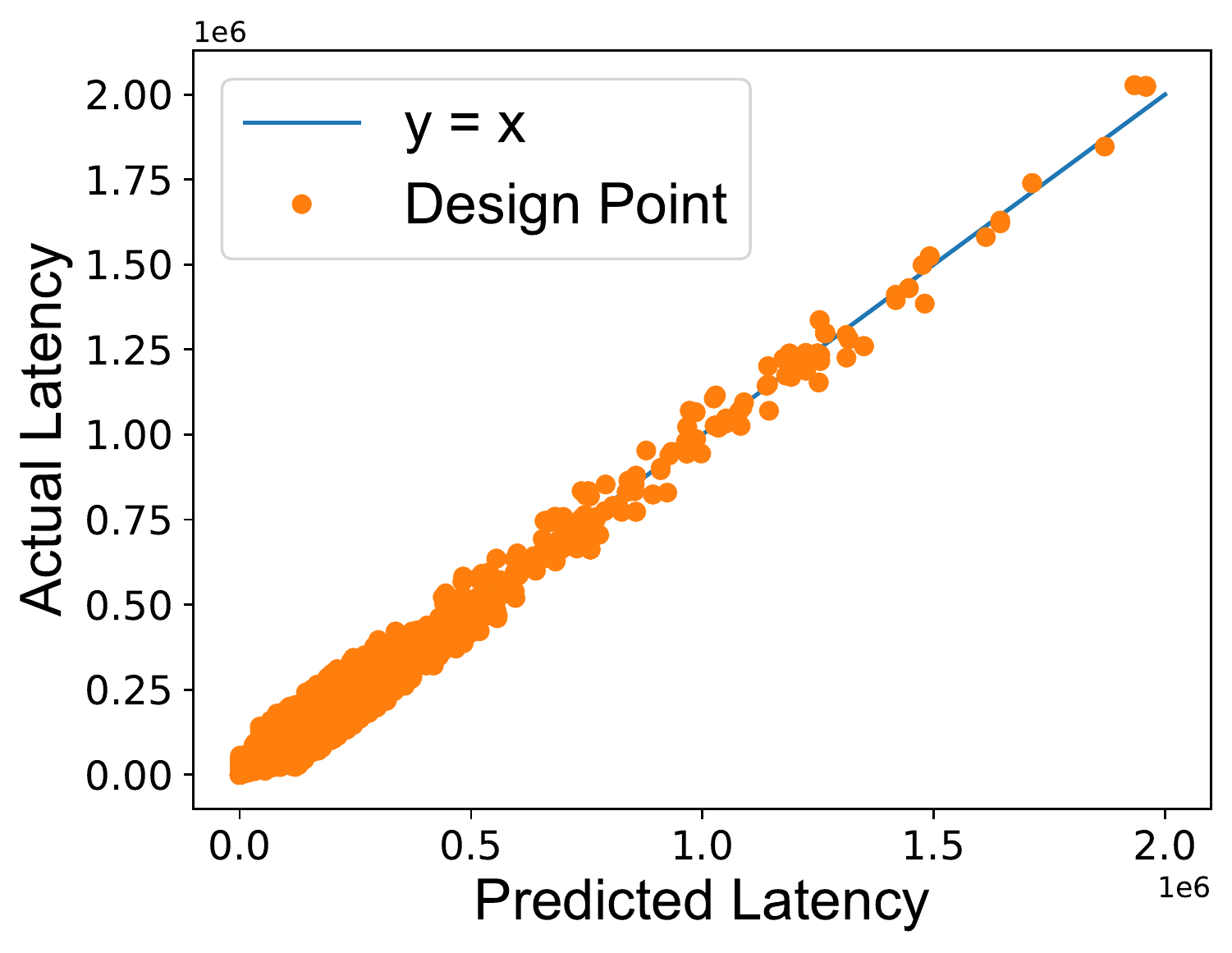}
    } 
    \subfigure[Prediction error with design size.]{
    \label{fig:b}
    \includegraphics[width= 0.46\columnwidth]{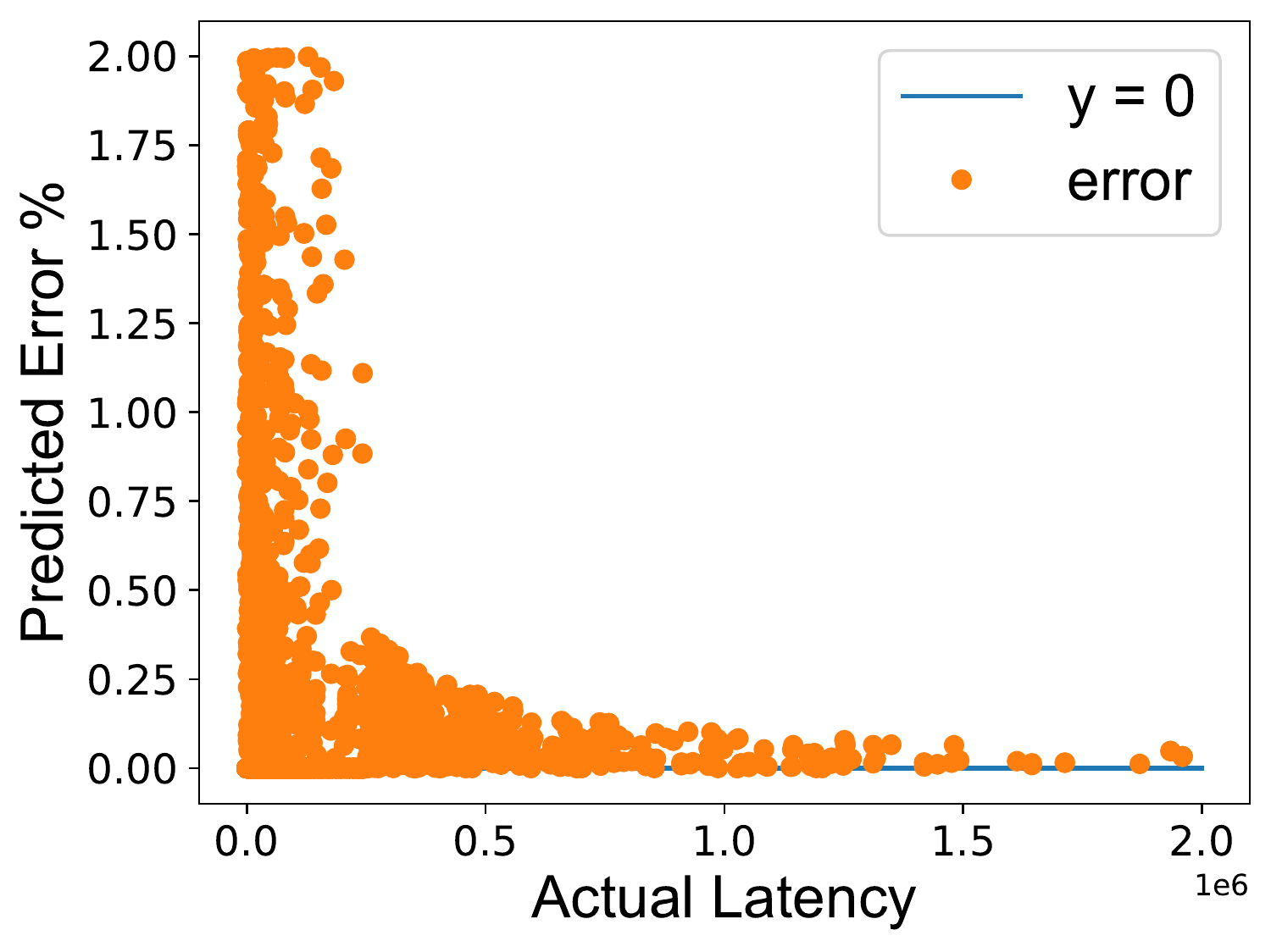}
    }
    \caption{Experiments of Latency model. (a) shows the matching of predicted and measured latency. (b) reveals the trend of prediction error w.r.t design size. The layer in larger workload has higher latency and smaller prediction error.}
    \label{fig: prediction model acurracy}
\end{figure}

As \archnames performs the DNN inference in a layer-wise synchronization manner, the total computation latency of target DNN with $N$-layers can be expressed as:
\begin{equation}
\label{eqt:latency_of_n3hcore}
  \mathrm{Latency}=\sum_{i=1}^N{\max \left( L_{\mathrm{LUT}}^{i},L_{\mathrm{DSP}}^{i} \right)}
\end{equation}
where the $L_{\mathrm{LUT}}^{i}$ and $L_{\mathrm{DSP}}^{i}$ denotes the computation latency taken by the LUT-core and DSP-core to complete the assigned workload for $i$-th layer respectively.
Since the instructions of our LUT-core and DSP-core are scheduled in static pipeline, the latency model formulated via accumulated execution time of schedule instruction is relatively accurate.
To verify our latency model, we test 10000 design points with design knobs listed on~\cref{table:params_to_optimize} of original manuscript using randomly generated value, where the result is shown in~\cref{fig: prediction model acurracy} with prediction error less than 2\%. 

\section{Hybrid Quantization with Mixed Precision}
\begin{figure}[t]
  \centering
  \includegraphics[width=.95\columnwidth]{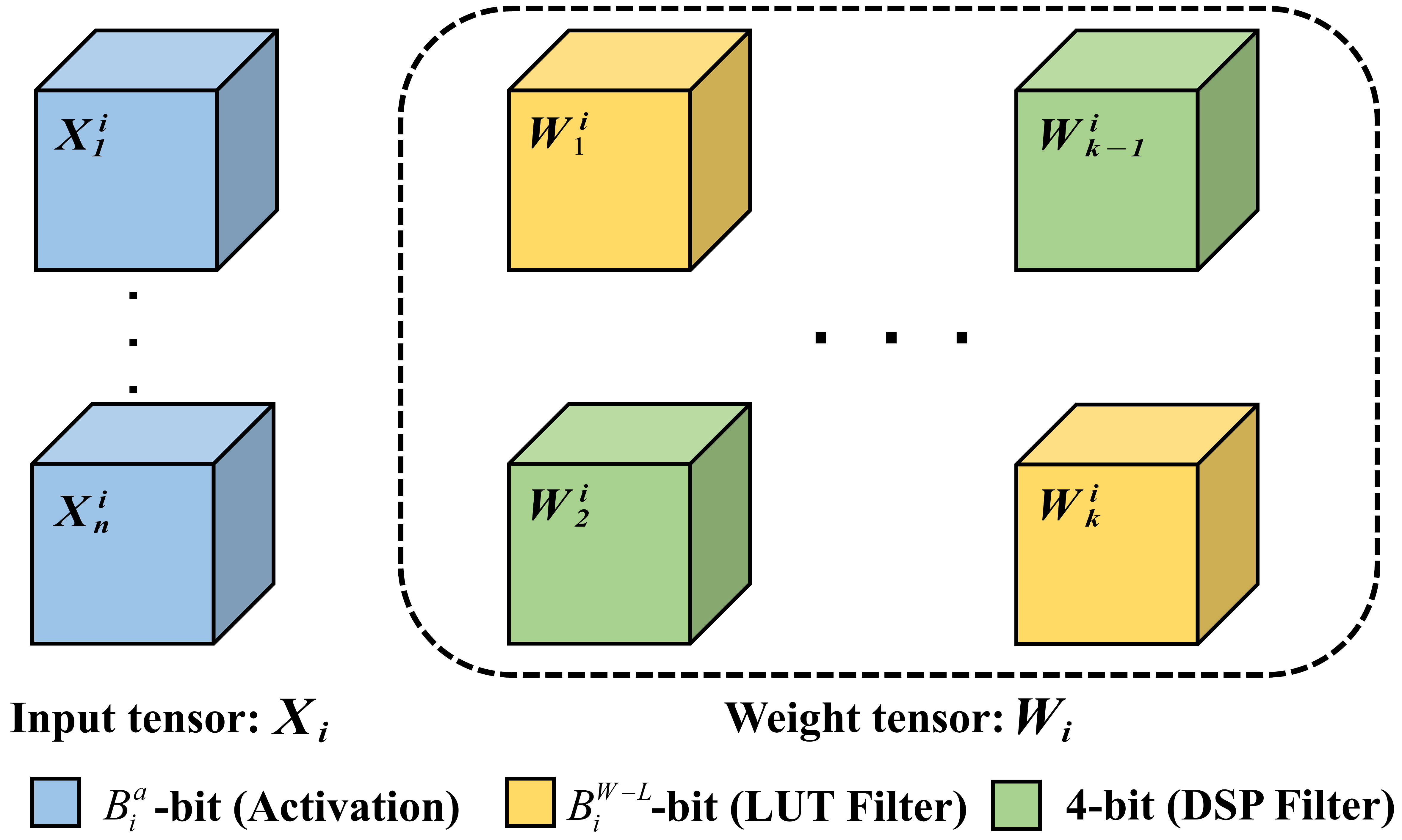}
  \caption{Proposed hybrid quantization scheme. Filters in each layer computed by DSP-core and LUT-core are quantized with uniform- and mixed-precision respectively.}
  \label{quantization}
\end{figure}

Suppose an $N$-layer DNN as $\left\{ W_i, X_i \right\} _{i=1}^{N}, W_i \in \mathbb{R}^{c_{\textrm{out}}^i\times c_{\textrm{in}}^i\times d^i\times d^i} $ and $X_i$ denotes the $i$-th layer weight and input/activation tensor respectively, where $d$ is the kernel size\footnote{$d=1$ for fully-connected layer}, $c_{\textrm{in}}$ and $c_{\textrm{out}}$ are number of input and output channels. 
The weight tensor $W_i$ can be viewed as a set $W_i=\left\{ W_{k}^{i} \right\} _{k=1}^{c_{\textrm{out}}^i}$ 
, where $W_{k}^{i}\in \mathbb{R}^{c_{in}^{i}\times d_{k}^{i}\times d_{k}^{i}}$ denotes the $k$-th filter. 
Our work applies filter-wise quantization, which takes a finer-grained strategy with $k$ as its granularity. 
For different filters in one layer, 3-D tensor $W_{k}^{i}$ is assigned with different quantization bit-width.

The hybrid quantization method integrates both uniform- and mixed-precision quantization in this work. 
The computation of one layer for DSP- and LUT-core is allocated in the granularity of filters. 
All the computations conducted by DSP-core are 4-bits quantized on weights.
The rest of filters computed by LUT-core are quantized with $B^{\textrm{w-L}}$ bits~($2\sim8$-bits) which is determined by our optimization framework~(\cref{sec: framwork}). 
Note that, $B^{\textrm{w-L}}$ varies in different layers. 
The filter allocation ratio will be discussed in~\cref{5.3}, which determines the number of filters allocated to DSP-/LUT-core on each layer. 
Besides, we need to consider which filters should be allocated to DSP-/LUT-core. 
In this work, the filter allocation~(to DSP/LUT) depends on the KL-divergence $\mathcal{D}_{\textrm{KL}}$ of the original filter weight distribution $W_{k}^{i}$ and its quantized weight distribution~(calibrated via one batch of images).
The filters with greater $\mathcal{D}_{\textrm{KL}}$ in one layer will be allocated to the computing core with higher bit-width, since the quantization bit-widths $B^{\textrm{w-L}}$ of LUT filters is flexible. 
Now, the filter number and type allocation problems are settled.

We adopt the layer-wise quantization manner for activations shared by both LUT- and DSP-core. 
The activation and weight of first and last layer are quantized into 8-bits.
In our work, we set the activation bit-widths range as 2-4 bits for all other layers, as the LUT-core enjoys much less latency with operands in low bit-width.  
Moreover, we linearly quantize the weights and activations of each layer using $ \left\{ 4,B^{\textrm{w-L}} \right\} $-bits and $\left\{B^{\textrm{a}} \right\}$-bits respectively. The linearly quantized model only requires fixed-point computing unit, which is more efficient in our heterogeneous architecture. The proposed quantization method of the $i^\mathrm{th}$ convolution layer is depicted in~\cref{quantization}. 

\section{Framework for Optimized System}
\label{sec: framwork}

\subsection{Framework Overview}
This section describes our framework to automatically generate the optimized system configuration for \archnames to achieve the optimized performance across varying DNNs and FPGAs.
The objects to be optimized by the framework come in threefold: 
\textit{1) Resource Allocation.} The framework configures the design parameters listed in~\cref{table:params_to_optimize} to achieve the optimized resource allocation for both LUT-core and DSP-core;
\textit{2) Workload Split.} 
As the \archnames splits the computing task into LUT- and DSP-core in the layer-wise fashion, the framework is expected to split the computation task for workload balance properly;
\textit{3) Quantization bit-width.} 
Since LUT-core supports operands with flexible bit-width while DSP-core only supports the fixed one~({\tt int4} in this work), the quantization bit-width affects the overall inference latency and accuracy.

\subsection{Resource allocation \& Quantization}
As discussed in~\cref{sec:arch}, we formulate the latency model for LUT- and DSP-core, and design factors influencing the latency and the corresponding ranges are tabulated in~\cref{range}. 
Design factors $K,M,N,D_{\textrm{L,buf}}^{\textrm{a}},$
$D_{\textrm{D,buf}}^{\textrm{a}},D_{\textrm{D,buf}}^{\textrm{w}}$ are related to the resource allocation. The left design factors $\left\{ B_i^{\textrm{w-L}},B_i^{\textrm{a}} \right\} _{i=1}^{N}$ are the quantization bit-widths of LUT filters and activations of each DNN layer. 
We leverage the Reinforcement Learning (RL) with DDPG algorithm~\cite{qiu2019deep} for the design space exploration~(can be viewed as sequential decision making) to identify the optimal system configuration with low latency but high accuracy.  
The change of design factors leads to different resource usage estimated by our cost model, which is constrained by the available resources on the target FPGA devices.

\begin{table}[t]
\caption{RL search range of design factors~(variables). \{D\textsubscript{A}, D\textsubscript{B}\} denotes the device \{XC7Z020, XC7Z045\}.}
\label{range}
\begin{tabular}{@{}lll@{}}
\toprule
Design Factors & Range on D\textsubscript{A} & Range on D\textsubscript{B} \\ \midrule
$K$              & $64\cdot v$   &  $64\cdot v$\\
               & $\left( 0<v<5, v\in \mathbb{N} \right)$ & $\left( 0<v<5, v\in \mathbb{N} \right)$       \\
$M$              & 1-50  & 1-252\\
$N$              & 1-50  & 1-252\\
$D_{\textrm{L,buf}}^{\textrm{a}}$             & $1024\cdot v$  & $1024\cdot v$ \\
               & $\left( 0<v\leqslant 50, v\in \mathbb{N} \right)$ & $\left( 0<v\leqslant 252, v\in \mathbb{N} \right)$ \\
$D_{\textrm{D,buf}}^{\textrm{a}}$           & $1024\cdot v$  & $1024\cdot v$\\
               & $\left( 0<v\leqslant 25, v\in \mathbb{N} \right)$  & $\left( 0<v\leqslant 126, v\in \mathbb{N} \right)$ \\
$D_{\textrm{D,buf}}^{\textrm{w}}$           & $1024\cdot v$  & $1024\cdot v$\\
               & $\left( 0<v<5, v\in \mathbb{N} \right)$  &  $\left( 0<v\leqslant 16, v\in \mathbb{N} \right)$ \\   \midrule
$\left\{ B_i^{\textrm{w-L}} \right\} _{i=1}^{N}$              & 2-8   & 2-8 \\
$\left\{ B_i^{\textrm{a}} \right\} _{i=1}^{N}$              & 2-4   & 2-4 \\ \bottomrule
\end{tabular}
\end{table}

\subsection{Neuron-based workload split}
\label{5.3}
\begin{figure}[t]
  \centering
  \includegraphics[width=0.7\columnwidth]{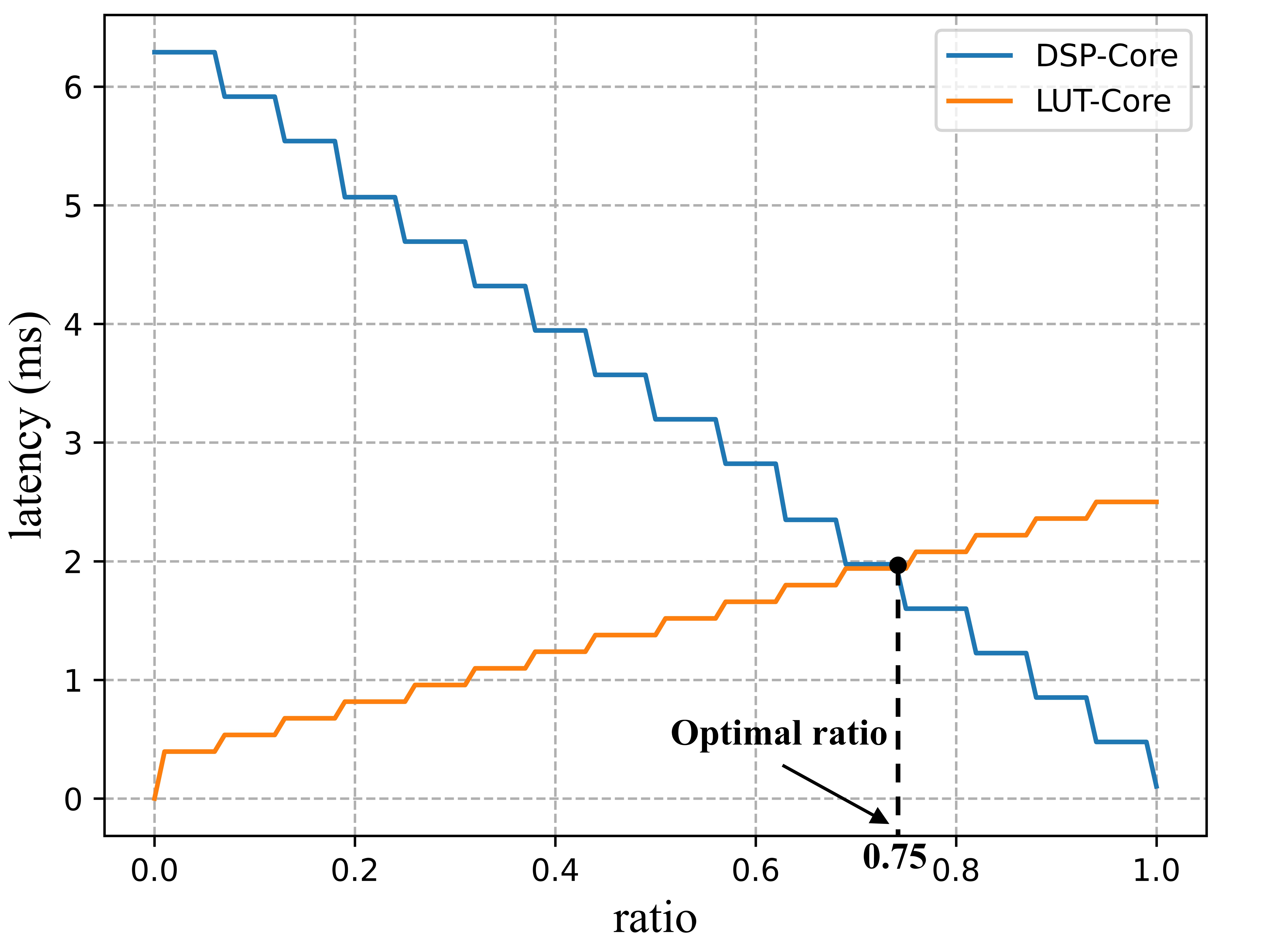}
  \caption{The relationship between latency and ratio of the $14$-th layer of ResNet18 with manual configuration (4 bits).}
  \label{lat}
\end{figure}

For each layer, the filter-based workload split ratio is defined as:
\begin{equation}
  \mathrm{ratio}_{i}=\frac{\mathrm{Filter}_{\mathrm{LUT}}^{i}}{\mathrm{Filter}_{\mathrm{all}}^{i}} 
\label{ratio}
\end{equation}
Where $\mathrm{Filter}_{\mathrm{all}}$ is the number of neurons/filters and $\mathrm{Filter}_{\mathrm{LUT}}$ is the number of filters computed by LUT-core.
After the quantization bit-widths $\left\{ B_i^{\textrm{w-L}},B_i^{\textrm{a}} \right\}$ of the $i$-th layer are determined by the agent, the best workload split ratio $\mathrm{ratio}_i$ is computed by minimizing the latency of the $i$-th layer, as defined:
\begin{equation}
\label{ratio2}
\begin{split}
    \underset{\mathrm{ratio}_i}{\argmin}\left( \max \left( L_\mathrm{LUT}^{i}\left( M,K,N,D_{\textrm{L,buf}}^{\textrm{a}},B_i^{\textrm{w-L}},B_i^{\textrm{a}}, \mathrm{ratio}_i \right), \right. \right. 
    \\
    \left. \left.  L_\mathrm{DSP}^{i}\left( D_{\textrm{D,buf}}^{\textrm{a}},D_{\textrm{D,buf}}^{\textrm{w-L}}, \mathrm{ratio}_i \right) \right) \right)
\end{split}
\end{equation}
The generated ratio is the optimal filter allocation strategy for inter-layer parallelism of \archname.
\cref{lat} shows an example of the relationship between the latency and ratio. From the figure, the optimized ratio is 0.75, which means $0.75\times256=192$ filters are computed by the LUT-core and the rest $256-192=64$ filters are assigned to DSP-core.

\subsection{RL implementation details}
\begin{figure}[h]
  \centering
  \includegraphics[width=.97\columnwidth]{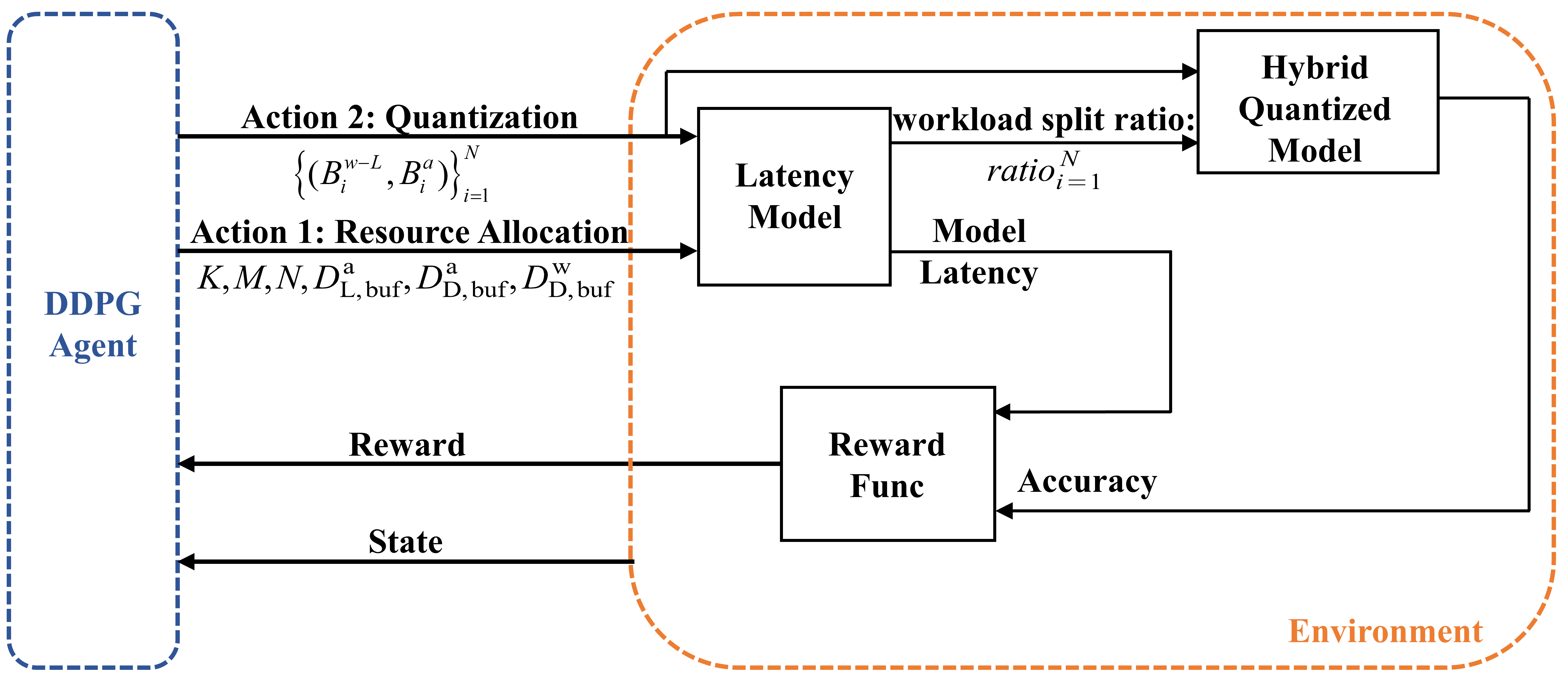}
  \caption{Overview of RL-based optimization framework.}
  \label{fig:RL-overview}
\end{figure}

The overview of our DDPG based RL learning framework~\cite{he2018amc} to explore the both the architecture configuration and workload assignment between heterogeneous cores is depicted in~\cref{fig:RL-overview}. Details of action space, state space, reward and etc are given as follows.

\subsubsection{Action Space.} 
The action space contains two tasks of actions: action-1 $a_1$ and action-2 $a_2$ in sequence. 
The action-1 decides the resource utilization design factors which is constrained by the available resources of given FPGA devices. \cref{range} provides the range of resource utilization design factors on two FPGAs.
In each episode, the RL agent takes 6+2N action steps, where N is the number of layers of given DNN. 

For action-1, the agent chooses the hardware-related factors $\left\{K,M,N,D_{\textrm{L,buf}}^{\textrm{a}},D_{\textrm{D,buf}}^{\textrm{a}},D_{\textrm{D,buf}}^{\textrm{w}} \right\} $
sequentially at the first 6 time steps. 
For each time-step, the agent output one action applies on one aforementioned hardware factor, which can be described as:
\begin{equation}
      a_1^t=\mathrm{round}\left( a_t \left( \mathrm{value}_{\max}- \mathrm{value}_{\min} \right) +\mathrm{value}_{\min} \right)
\end{equation}
where $t\in [0,5]$. The continuous action $a_t$ (range: [0, 1]) of the DDPG agent are discretized by round function$\mathrm{round}()$ into the range of 6 design factors in integer.
Note that, $K$,$D_{\textrm{L,buf}}^{\textrm{a}}$, $D_{\textrm{D,buf}}^{\textrm{a}}$, $D_{\textrm{D,buf}}^{\textrm{w}}$ are in the format of $c\cdot v$ where $c$ is a constant parameter and $v$ is a variable. 
For these 4 parameters, $\left[ \mathrm{value}_{\min}, \mathrm{value}_{\max} \right] $ is the range of the parameter $v$, so the actions $\left[ a_1^0,a_1^3,a_1^4,a_1^5 \right]$ will be multiplied with their corresponding constant parameters $c$ which are shown in~\cref{range}.

For action-2, the agent chooses the software-related factors $\{B^\textrm{w-L}, B^\textrm{a}\}$ with 2N steps. 
The actions can be denoted as:
\begin{equation}
 a_2^t=\textrm{round}(a_t\times (b_{\max}-b_{\min}+1)+b_{\min}-0.5) 
\end{equation}
where $[b_{\min}, b_{\max}]$ is the bit-width range in~\cref{range}.

\subsubsection{State Space}
The state space $S_t$ for the $t$-th time step is expressed as:
\begin{equation}
    S_t=\left( \mathrm{id}_{\mathrm{func}}, i, \mathrm{NN}_i, \mathrm{ratio}_i, a_t \right) 
\end{equation}
where $\mathrm{id}_{\mathrm{func}}$ is the binary indicator for the current task (0 for $a_1^t$, 1 for $a_2^t$) , $i$ is the layer index, $\mathrm{NN}_i$ is the DNN information of the $i$-th layer, $\mathrm{ratio}_i$ is the calculated filter number allocation ratio in the $i$-th layer, $a_t$ is the action of the current time step. $i$, $\mathrm{NN}_i$ and $\mathrm{ratio}_i$ are set as 0 when $\mathrm{id}_{\mathrm{func}}$=0.
The DNN information of the $i$-th layer $\mathrm{NN}_i$ is a group of parameters defined as:
\begin{equation}
\mathrm{NN}_i = \left(c_{\mathrm{in}}^i,c_{\mathrm{out}}^i,s_{\mathrm{kernel}}^i,s_{\mathrm{stride}}^i,s_{\mathrm{fmap}}^i,n_{\mathrm{params}}^i,\mathrm{id}_{\mathrm{sc}\&\mathrm{dw}},\mathrm{id}_{\mathrm{m}/\mathrm{n}}\right)
\end{equation}
where $c_{\mathrm{in}}$ and $c_{\mathrm{out}}$ are \#input/output channels, $s_{\mathrm{kernel}}, s_{\mathrm{stride}}$  and $s_{\mathrm{fmap}}$ are the sizes of the kernel, stride and feature map. $n_{\mathrm{params}}$ denotes \#parameters. 
$\mathrm{id}_{\mathrm{sc}\&\mathrm{dw}}$ is the binary index for shortcut layers of ResNet~\cite{he2016deep} and depthwise layers of MobileNet~\cite{sandler2018mobilenetv2}. 
If shortcut or depthwise, it is set 1, otherwise 0. 
$\mathrm{id}_{\mathrm{m/n}}$ is the binary index for the type of the current action: 0 for quantization of LUT filters $B_i^{\textrm{m-L}}$, 1 for quantization of activations $B_i^{\textrm{a}}$.

Thus, the entire state space $S_t$ can be unfolded as:
\begin{equation}
\begin{split}
    S_t=
    \left( \mathrm{id}_{\mathrm{func}},i,c_{\mathrm{in}}^i, c_{\mathrm{out}}^i, s_{\mathrm{kernel}}^i, s_{\mathrm{stride}}^i, s_{\mathrm{fmap}}^i, n_{\mathrm{params}}^i, \right.  \\ 
    \left. \mathrm{id}_{\mathrm{sc}\&\mathrm{dw}}, \mathrm{id}_{\mathrm{m}/\mathrm{n}}, \mathrm{ratio}_i,a_t \right)
\end{split}
\end{equation}
In addition, we normalize each dimension of the state space $S_t$ into [0, 1] to make them in the same scale.

\subsubsection{Reward.} After actions are taken in each episode, the model latency with current design configuration is estimated by our latency model, and the DNN is quantized w.r.t the identified quantization scheme. 
Then, we retrain the quantized model for one epoch to recover the accuracy, denoted as $\mathrm{acc}_{\mathrm{q}}$.
The reward function $\mathcal{R}$ is designed by combining the model latency $L_{\mathrm{m}}$ and accuracy $\mathrm{acc}_{\mathrm{q}}$, which is formulated as:
\begin{equation}
\mathcal{R}=\begin{cases}
	\frac{L_{\mathrm{t}}-L_{\mathrm{m}}}{L_{\mathrm{t}}}-1&		L_{\mathrm{m}}>L_{\mathrm{t}}\\
	\left( \mathrm{acc}_{\mathrm{q}}- \mathrm{acc}_{\mathrm{b}} \right) \cdot \lambda&		L_{\mathrm{m}}\leqslant L_{\mathrm{t}}\\
\end{cases}
\end{equation}
where $L_{\mathrm{t}}$ is the latency bound or target latency, $\mathrm{acc}_{\mathrm{b}}$ is the baseline accuracy of our model, $\lambda=0.01$ is a scaling factor. 
Note that, the reward remains less than -1 when the model latency exceeds the latency bound, otherwise the reward is in the interval of (-1, 1).


\section{Experiments}
In this section, we conduct experiments on various representative DNNs and FPGA devices to demonstrate the merits of hardware and software parts of \archname. 

\subsection{Experimental Setup}
\label{6.1}

\textbf{Networks and Dataset.} 
In this work, we uses two the classic image classification neural architectures, i.e., ResNet-18~\cite{he2016deep} and MobileNet-V2~\cite{sandler2018mobilenetv2} with large-scale ImageNet dataset~\cite{deng2009imagenet}. The data augmentation is identical as stated in~\cite{he2016deep}. The full-precision pretrained model utilized in the \archnames is obtained from the Pytorch model hub~\cite{paszke2017automatic}.
\textbf{Hyper-parameters of software framework.} The RL agent explores 900 episodes for each experiment setting. For each episode, if the model latency is within the latency bound, we will retrain the quantized model for one epoch to recover its accuracy for reward evaluation. 
We use SGD as optimizer with a fixed learning rate of 0.001 and momentum of 0.9. 
We randomly sample 100 categories of ImageNet to accelerate the exploration process.  
\textbf{FPGA platforms.}
Two FPGA boards, i.e., XC7Z020 and XC7Z045, are chosen as the target hardware platforms. 
With the \archnames architecture deployed upon different FPGAs, the device-specific configuration scheme is given by the framework automatically, constrained by the available hardware resources~(e.g., DSP, LUT and BRAM). 
The operating frequency is set to 100MHz.

\begin{table}[t]
\caption{Architecture design configuration (Config.) of \archnames automatically generated by the framework. \{D, N, T\} denotes \{Device, Neural Network, Target latency\} respectively. \{D\textsubscript{A
}, D\textsubscript{B}\} denotes the device \{XC7Z020, XC7Z045\}, and \{N\textsubscript{ResNet}, N\textsubscript{MobileNet}\} denotes the target DNN of \{ResNet18, MobileNet-v2\}, and T\textsubscript{$t$ms} denotes the target latency set in the framework is $t$-ms ($t\in \{5,6,7,25,30,35\}$). }
\label{conifgs}
\resizebox{\columnwidth}{!}{
\begin{tabular}{@{}c|cccccc@{}}
\toprule
          Configuration & K   & M & N  & $D_{\textrm{L,buf}}^{\textrm{a}}$   & $D_{\textrm{D,buf}}^{\textrm{a}}$ & $D_{\textrm{D,buf}}^{\textrm{w}}$ \\ \midrule
 D\textsubscript{A}N\textsubscript{ResNet}T\textsubscript{30ms} & 128 & 7 & 17 & 1024 & 1024$\cdot$2 & 1024 \\
 D\textsubscript{A}N\textsubscript{ResNet}T\textsubscript{35ms} & 128 & 8 & 16 & 1024 & 1024$\cdot$2 & 1024 \\ 
 D\textsubscript{A}N\textsubscript{MobileNet}T\textsubscript{5ms} & 64 & 18 & 12 & 1024 & 1024$\cdot$11 & 1024 \\
 D\textsubscript{A}N\textsubscript{MobileNet}T\textsubscript{7ms} & 64 & 26 & 8 & 1024 & 1024$\cdot$9 & 1024 \\ \midrule
 D\textsubscript{B}N\textsubscript{ResNet}T\textsubscript{25ms} & 512 & 11 & 17 & 1024 & 1024$\cdot$6 & 1024$\cdot$2 \\
 D\textsubscript{B}N\textsubscript{ResNet}T\textsubscript{30ms} & 512 & 14 & 14 & 1024 & 1024$\cdot$15 & 1024 \\
 D\textsubscript{B}N\textsubscript{MobileNet}T\textsubscript{5ms} & 64 & 35 & 22 & 1024 & 1024$\cdot$19 & 1024$\cdot$11 \\
 D\textsubscript{B}N\textsubscript{MobileNet}T\textsubscript{6ms} & 64 & 44 & 18 & 1024 & 1024$\cdot$20 & 1024$\cdot$8 \\ 
\bottomrule
\end{tabular}}
\end{table}

\subsection{Results and Analysis}
\begin{figure}[t]
  \centering
  \includegraphics[width=\columnwidth]{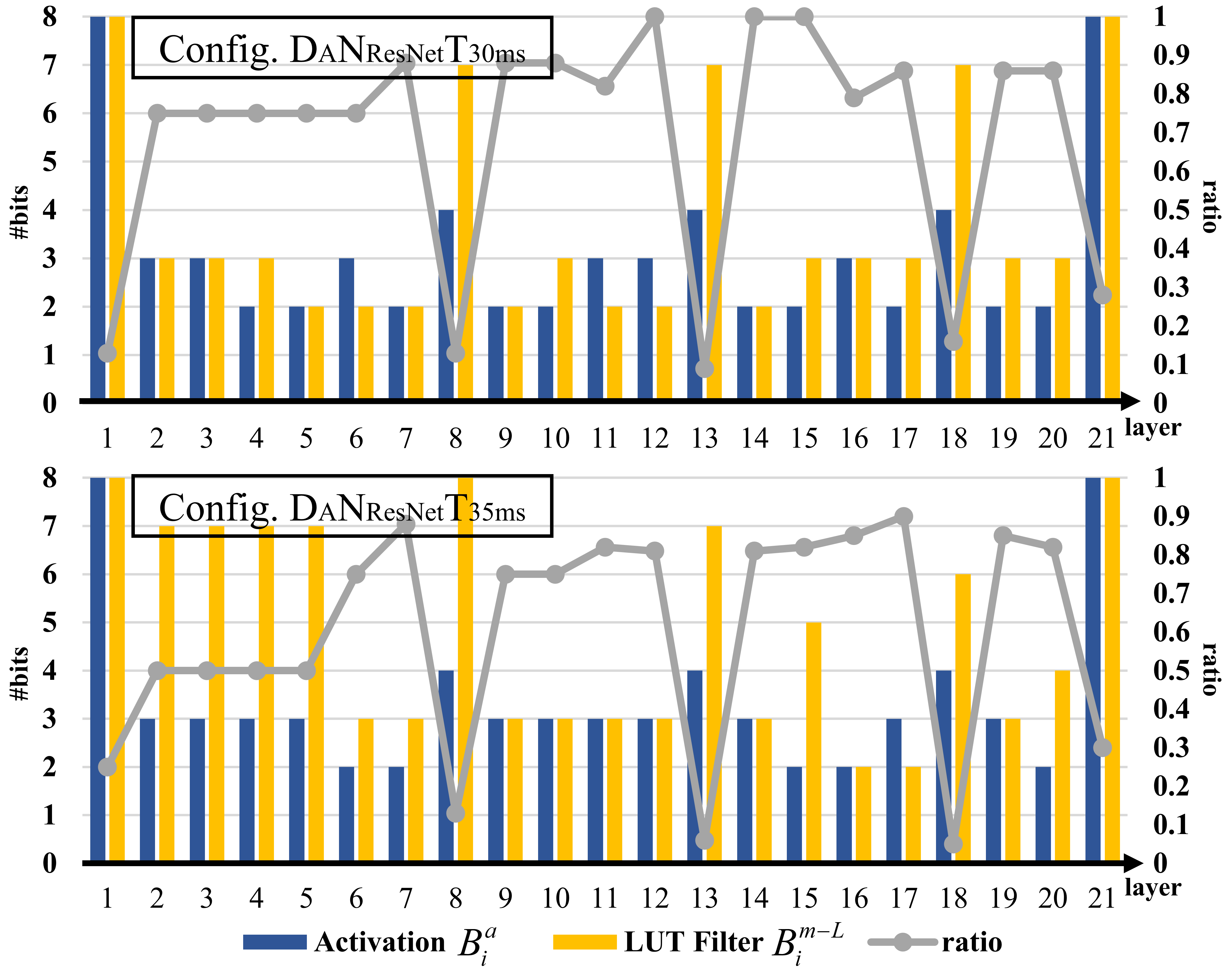}
  \caption{Layer-wise bit-width setting and workload split ratio in Config. D\textsubscript{A}N\textsubscript{ResNet}T\textsubscript{30ms} and D\textsubscript{A}N\textsubscript{ResNet}T\textsubscript{35ms}.}
  \label{DA}
\end{figure}

\begin{figure}[t]
  \centering
  \includegraphics[width=\columnwidth]{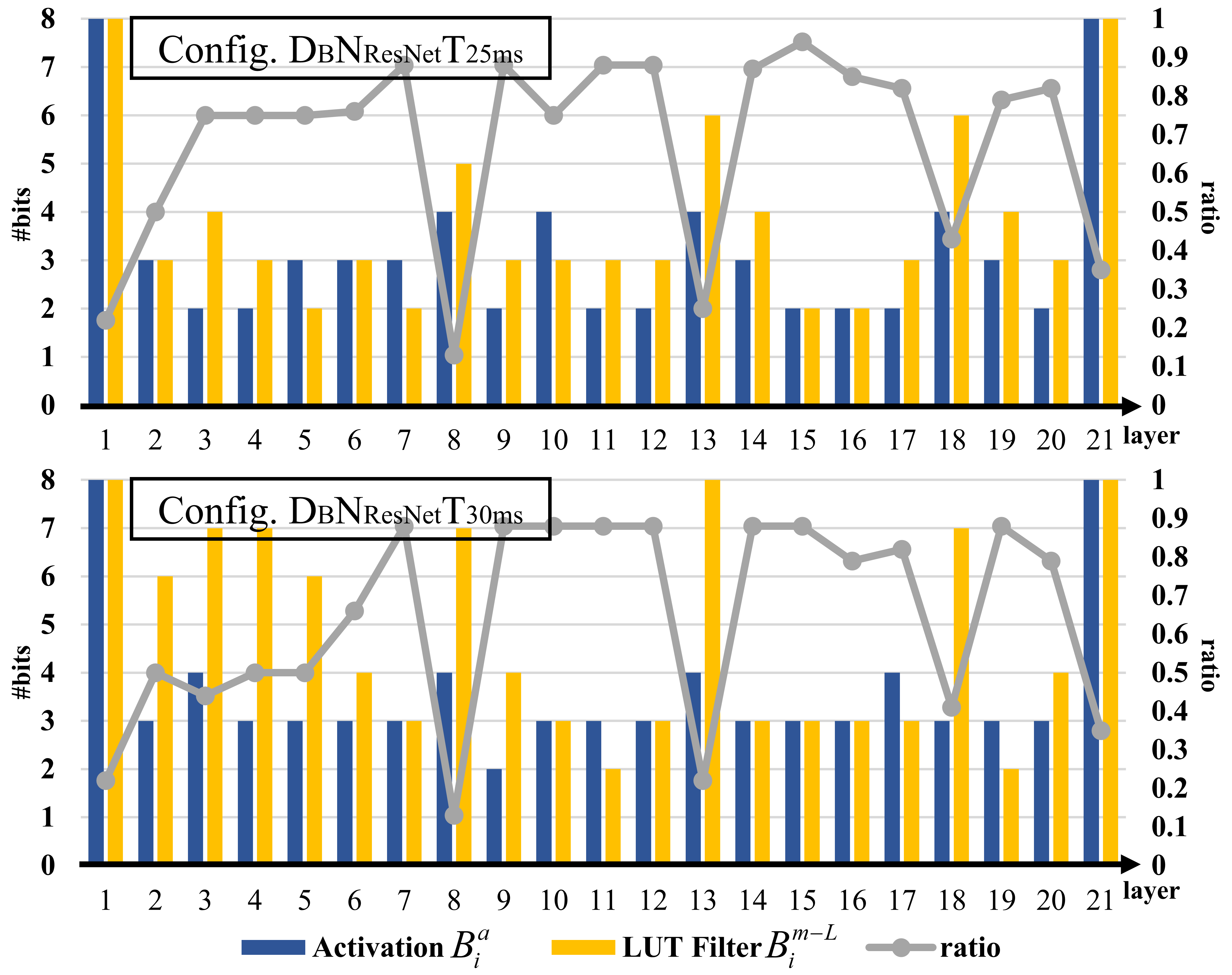}
  \caption{Layer-wise bit-width setting and workload split ratio in Config. of D\textsubscript{B}N\textsubscript{ResNet}T\textsubscript{25ms} and D\textsubscript{B}N\textsubscript{ResNet}T\textsubscript{30ms}.}
  \label{DB}
\end{figure}

\begin{figure}[t]
  \centering
  \includegraphics[width=\columnwidth]{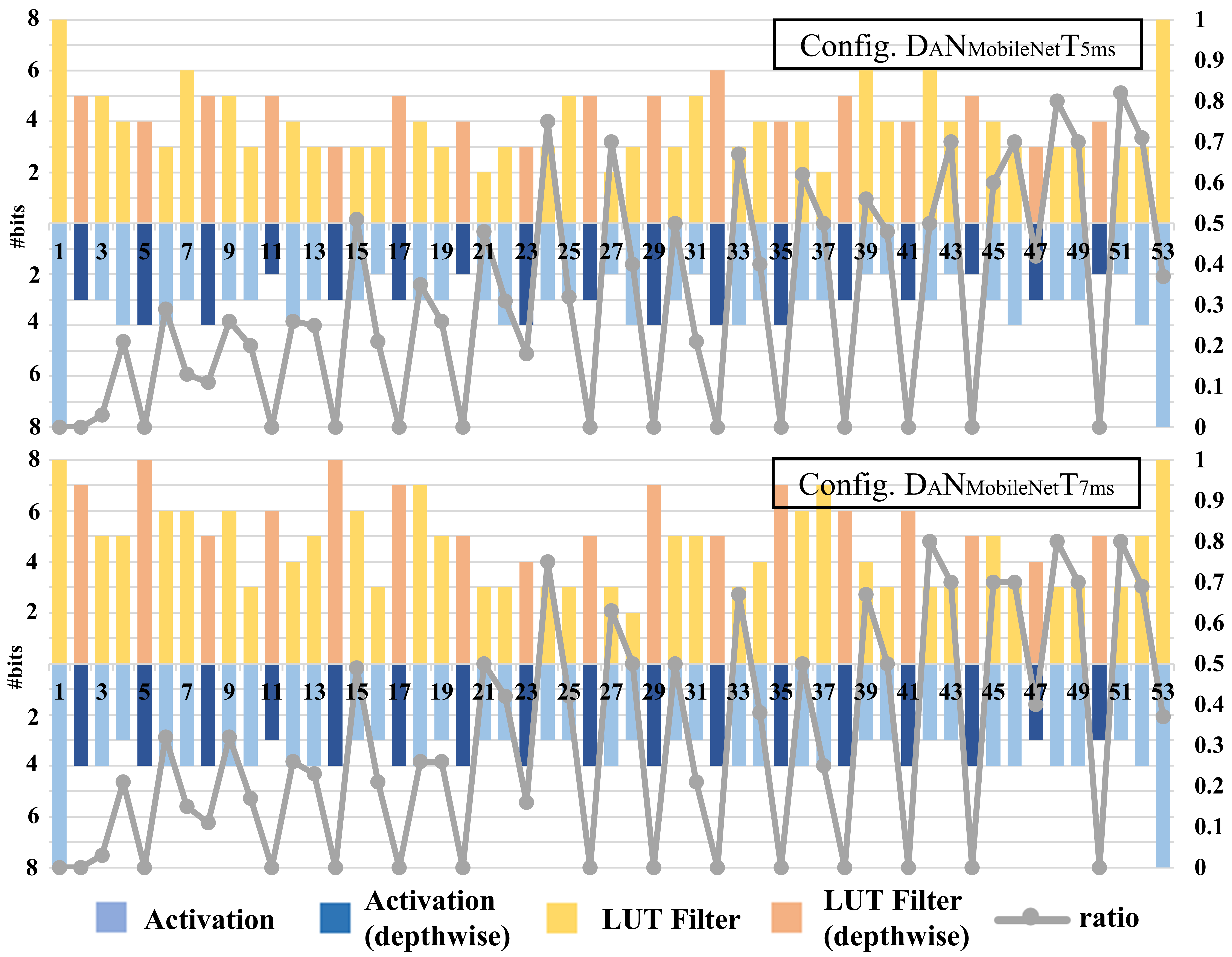}
  \caption{Layer-wise bit-width setting and workload split ratio in Config. D\textsubscript{A}N\textsubscript{MobileNet}T\textsubscript{5ms} and D\textsubscript{A}N\textsubscript{MobileNet}T\textsubscript{7ms}.}
  \label{DA2}
\end{figure}

\begin{figure}[t]
  \centering
  \includegraphics[width=\columnwidth]{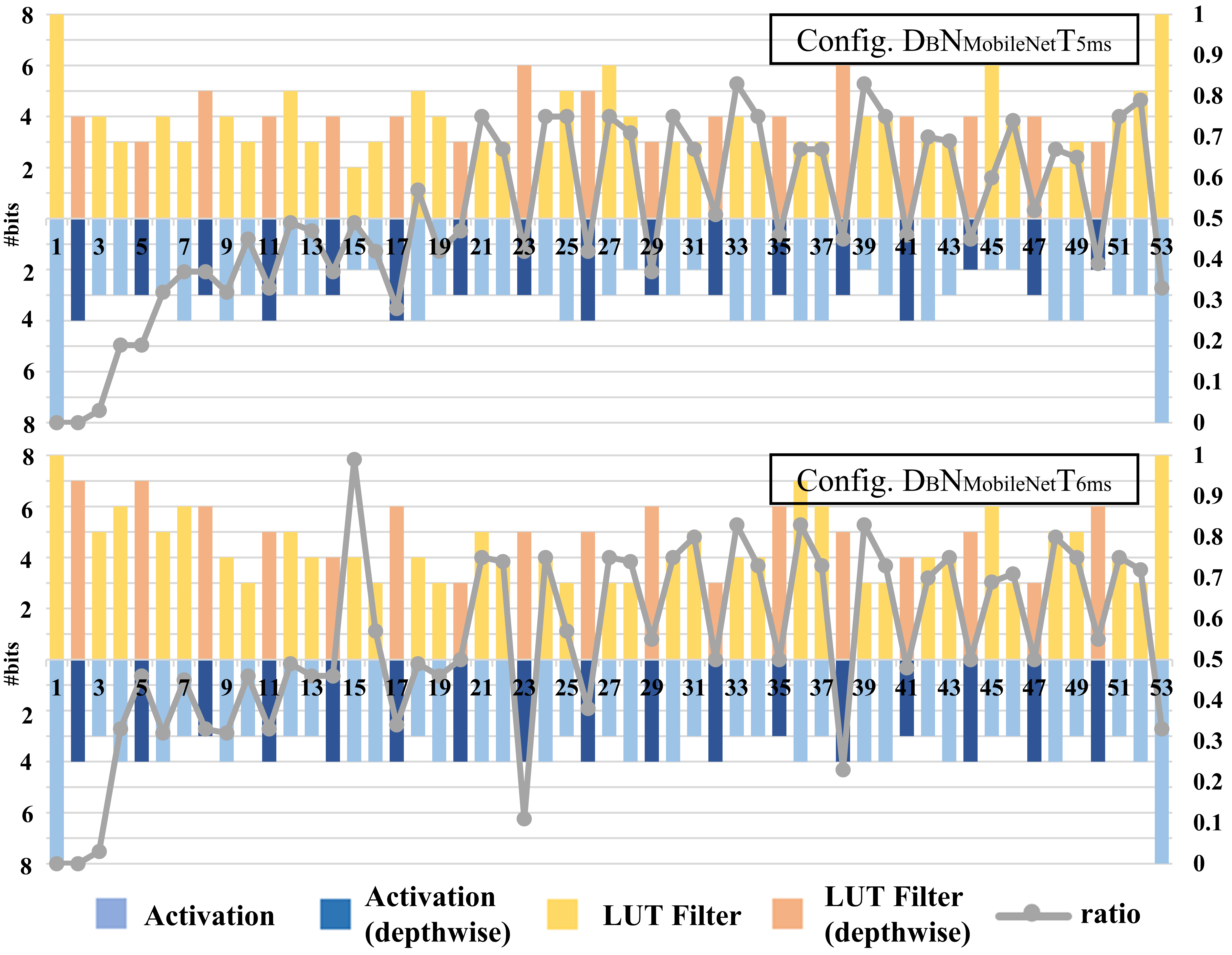}
  \caption{Layer-wise bit-width setting and workload split ratio in Config. D\textsubscript{B}N\textsubscript{MobileNet}T\textsubscript{5ms} and D\textsubscript{B}N\textsubscript{MobileNet}T\textsubscript{6ms}. }
  \label{DB3}
\end{figure}

We evaluate the framework under different target latency bounds, and generate 8 configuration combinations for the two DNNs and two FPGAs. 
The accuracy and latency of each configuration are tabulated in~\cref{acc_table}, where the configuration details are given in~\cref{conifgs}. 
For each configuration, we also provide the quantization bit-widths and the workload-split ratio of each layer in~\cref{DA,DB,DA2,DB3}. 

\begin{table*}[t]
\caption{Comparisons with state-of-the-art implementations on ImageNet}
\label{compare}
\begin{threeparttable}[b]
\scalebox{0.8}{
\begin{tabular}{@{}c|cc|cc|cc|cc|cc@{}}
\toprule
Implementation      & \multicolumn{2}{c|}{\begin{tabular}[c]{@{}c@{}}MobileNet-V2\\ \cite{wu2019high}\end{tabular}} & \multicolumn{2}{c|}{\begin{tabular}[c]{@{}c@{}}ResNet-18\\ \cite{9407035}\end{tabular}} & \multicolumn{2}{c|}{\begin{tabular}[c]{@{}c@{}}MobileNet-V2\\ \cite{9407035}\end{tabular}}  & \multicolumn{2}{c|}{\begin{tabular}[c]{@{}c@{}}ResNet-18\\ Ours\tnote{1} \end{tabular}} & \multicolumn{2}{c}{\begin{tabular}[c]{@{}c@{}}MobileNet-V2\\ Ours\tnote{2} \end{tabular}} \\ \midrule
Device              & ZU2EG          & ZU9EG           & XC7Z020       & XC7Z045       & XC7Z020         & XC7Z045        & XC7Z020                                         & XC7Z045                                         & XC7Z020                                           & XC7Z045                                          \\ \midrule
Bit-width (W/A)     & \multicolumn{2}{c|}{8/8}          & \multicolumn{2}{c|}{4/4}       & \multicolumn{2}{c|}{4/4}          & Flexible                                        & Flexible                                        & Flexible                                          & Flexible                                         \\ \midrule
Top-1 Accuracy (\%) & \multicolumn{2}{c|}{68.1}         & \multicolumn{2}{c|}{70.27}     & \multicolumn{2}{c|}{65.64}        & 70.45                                           & 70.39                                           & 66.25                                             & 66.04                                            \\ \midrule
Frequency (MHz)     & 430            & 333             & \multicolumn{2}{c|}{100}       & \multicolumn{2}{c|}{100}          & \multicolumn{2}{c|}{100}                                                                           & \multicolumn{2}{c}{100}                                                                              \\ \midrule
LUT                 & 31198          & 161944          & 28288         & 145049        & 28288           & 145049         & 39623                                           & 152868                                          & 45765                                             & 192624                                           \\
DSP                 & 212            & 2070            & 220           & 900           & 220             & 900            & 220                                             & 900                                             & 220                                               & 900                                              \\
BRAM                & 145            & 771             & 56            & 225.5         & 56              & 225.5          & 137                                             & 541                                             & 137                                               & 461                                              \\ \midrule
Latency (ms)        & -              & -               & 47.10         & 40.36         & 8.29            & 7.28           & 35.79                                           & 32.47                                           & 7.51                                              & 6.62                                             \\
Throughput (GOPS)   & -              & -               & 77.0          & 359.2         & 71.8            & 326.9          & 101.3                                           & 446.8                                           & 80.1                                              & 363.5                                            \\
Frame Rate (FPS)    & 205.3          & 809.8           & 21.3          & 99.1          & 120.7           & 549.3          & 27.9                                            & 123.2                                           & 133.2                                             & 604.2                                            \\
GOPS/DSP            & -              & -               & 0.350         & 0.391         & 0.326           & 0.363          & 0.460                                           & 0.496                                           & 0.364                                             & 0.404                                            \\
GOPS/kLUT           & -              & -               & 2.725         & 2.475         & 2.538           & 2.252          & 2.557                                           & 2.923                                           & 1.750                                             & 1.887                                            \\ \bottomrule
\end{tabular}
}
    \begin{tablenotes}
    \footnotesize
     \item[1] Config. D\textsubscript{A}N\textsubscript{ResNet}T\textsubscript{35ms} \& Config. D\textsubscript{B}N\textsubscript{ResNet}T\textsubscript{30ms}
     \item[2] Config. D\textsubscript{A}N\textsubscript{MobileNet}T\textsubscript{7ms} \& Config. D\textsubscript{B}N\textsubscript{MobileNet}T\textsubscript{6ms}
    \end{tablenotes}
    
\end{threeparttable}
\end{table*}

\begin{table}[t]
\centering
\caption{Evaluation Accuracy, IDE simulated and FPGA measured latency of \archnames running with ResNet18 and MobileNet-v2 on ImageNet dataset.}
\label{acc_table}
\resizebox{\columnwidth}{!}{
\begin{tabular}{@{}ccccc@{}}
\toprule
\begin{tabular}[c]{@{}c@{}} \archnames \\ Configuration\end{tabular}
& \begin{tabular}[c]{@{}c@{}}Bit width\\ (W./A.)\end{tabular} & \begin{tabular}[c]{@{}c@{}}Accuracy (\%)\\ (Top1/Top5)\end{tabular} & \begin{tabular}[c]{@{}c@{}}Model\\ latency~(ms)\end{tabular} & \begin{tabular}[c]{@{}c@{}}Measured\\ latency~(ms)\end{tabular} \\ \midrule
\multicolumn{5}{c}{ResNet-18}                                                                                                                                                                                                                                                                                          \\ \midrule
Pretrained Baseline                                                                                    & 32/32                                                             & 69.76/89.08                                                         & N/A                                      & N/A                                     \\
Manual Config. D\textsubscript{A}N\textsubscript{ResNet}                                                                            & 4/4                                                               & 69.65/89.11                                                         & 40.96                                  & 42.26                                 \\
Auto. Config. D\textsubscript{A}N\textsubscript{ResNet}T\textsubscript{30ms}                                                                          & Flexible                                                          & 67.28/87.85                                                         & 29.14                                  & \textbf{31.43}                                 \\
Auto. Config. D\textsubscript{A}N\textsubscript{ResNet}T\textsubscript{35ms}                                                                         & Flexible                                                          & \textbf{70.45/89.54 }                                                        & 34.95                                  & 35.79                                 \\ \midrule
Manual Config. D\textsubscript{B}N\textsubscript{ResNet}                                                                             & 4/4                                                               & 69.65/89.11                                                         & 30.26                                  & 32.77                                 \\
Auto. Config. D\textsubscript{B}N\textsubscript{ResNet}T\textsubscript{25ms}                                                                          & Flexible                                                          & 66.81/87.19                                                         & 24.83                                  & \textbf{26.90}                                 \\
Auto. Config. D\textsubscript{B}N\textsubscript{ResNet}T\textsubscript{30ms}                                                                        & Flexible                                                          & \textbf{70.39/89.33  }                                                       & 29.52                                  & 32.47                                 \\ \midrule
\multicolumn{5}{c}{MobileNet-V2}                                                                                                                                                                                                                                                                                      \\ \midrule
Pretrained Baseline                                                                                    & 32/32                                                             & 71.88/90.29                                                         & N/A                                      & N/A                                     \\
Mannual Config. D\textsubscript{A}N\textsubscript{MobileNet}                                                                            & 4/4                                                               & 65.18/75.77                                     & 8.85                                   & 9.15                                  \\
Auto. Config. D\textsubscript{A}N\textsubscript{MobileNet}T\textsubscript{5ms}                              & Flexible                                                          & 62.76/73.32                                     & 4.93                                   & \textbf{5.66}                                  \\
Auto. Config. D\textsubscript{A}N\textsubscript{MobileNet}T\textsubscript{7ms}                                                                            & Flexible                                                          & \textbf{66.25/76.09}                                     & 6.95                                   & 7.51                                  \\ \midrule
Auto. Config. D\textsubscript{B}N\textsubscript{MobileNet}T\textsubscript{5ms}                                                                          & Flexible                                                          & 63.41/73.56                                     & 4.86                                   & \textbf{5.33}                                  \\
Auto. Config. D\textsubscript{B}N\textsubscript{MobileNet}T\textsubscript{6ms}                                                                            & Flexible                                                          & \textbf{66.04/75.88}                                     & 5.93                                   & 6.62                                  \\ \bottomrule
\end{tabular}
}
\end{table}

\subsubsection{Resource allocation Analysis}
From~\cref{conifgs}, on D\textsubscript{A} we can find that $K$ is greater when the DNN is ResNet-18. 
$K$ denotes the computational parallelism of input channels of activation. A large $K$ allows more input channels to be computed in parallel, improving the computation efficiency.
Compared with ResNet, MobileNet contains depth-wise layers which occupy one-third of the number of layers. 
Since depth-wise layers has a small number of input channels, the agent tends to assign a small $K$ to LUT-Core. 
It indicate that LUT-Core is more efficient in computing ResNet than MobileNet.
Thus, for MobileNet, the agent allocates more BRAMs to DSP-Core, which allows DSP-Core compute faster to compensate for the inefficiency of LUT-Core.
As shown in~\cref{DA2}, for the same reason, the split ratios of many depthwise layers in MobileNet are zero, which means that the workload of the depthwise layer is solely assigned to the DSP-Core for minimized latency.

\subsubsection{Quantization Analysis}
In~\cref{DA}, it reveals that the RL agent assigns more bit-width in down-sampling layers~(layer 8, 13, 18). 
To avoid the information bottleneck, higher bit-width are assigned to those layers~(e.g., 7-bits for LUT-Core filters and 4-bits for shared activations).
When the target latency bound is set to 30ms, the quantization bit-width becomes lower under the strict limitation. 
Consequently, the LUT-Core latency is reduced, and more workload is assigned to the LUT-Core to keep the latency below the bound. 
This proves that the agent can automatically analyze the latency and adjust the bit-width and ratio.
As shown in~\cref{DA2}, LUT-Core is not efficient to compute depth-wise layers, thus the workload split ratio of depth-wise layers is lower than point-wise layers. When the bound is set to 5ms, the bit-width across layers is further lowered compared to the 7ms counterpart.
On D\textsubscript{B}, the configuration of quantization and ratio shows the same trend.

\subsubsection{Comparison with the state-of-arts accelerators}
We selected the results of ResNet18 and MobileNet networks on two devices XC7Z020 and XC7Z045, as shown in~\cref{compare}, and compare the experimental results with the existing state-of-arts designs.
When deploying ResNet18 on the given two FPGA devices, we obtain comparable accuracy while reducing latency by $1.32\times$ and $1.24\times$ compared to the Mix\&Match design~\cite{9407035}. 
The Mix\&Match is also a heterogeneous architecture utilizing DSP and LUT resources, but without exploring the design space, our latency-centric architecture outperforms it on resource utilization, quantization bit-width flexibility, and performance~(e.g., latency, throughput and frame rate), even though they adopt more hardware-friendly power-of-2~\cite{liu2021improving} quantization scheme.

As can be found in~\cref{compare}, our proposed optimization framework makes better utilization of the on-chip resources, consuming more BRAMs and LUTs to boost the performance. 
In terms of the throughput per resource unit, the GOPS/DSP is higher on both devices. 
The GOPS/kLUT on XC7Z020 is less, but when more batches are computed simultaneously on XC7Z045, it outperforms the design in~\cite{9407035}.
On MobileNet, we get the speedup by about $1.12\times$ on both FPGA devices. The lower performance improvement on MobileNet over ResNet is due to the low computation efficiency of LUT-Core on MobileNet. This is also reflected in the efficiency of hardware resource usage. Despite the performance improvement, the throughput per resource unit on both devices is lower than Mix\&Match~\cite{9407035}. 
Therefore, the proposed heterogeneous architecture has a limited effect on depth-wise acceleration.
Moreover, in terms of the frame-rate~(FPS), our design outperforms the mix\&match, and will outperforms design in~\cite{wu2019high}~($2.48\sim2.89\times$ FPS) as well when it operates at 100MHz.


\section{Conclusion}
In this paper, we propose a heterogeneous architecture on FPGA with two cores~(DSP/LUT-core) computing synchronously for efficient DNN inference. 
We also propose a hybrid quantization scheme which quantizes the NN weights according to the filter computation allocation. 
To rapid hardware-mapping evaluation and find the optimal resource allocation~(BRAM, LUT) scheme for DSP- and LUT-core, we build the latency and cost models for both that parameterized by architecture configuration knobs. 
Based on aforementioned models, we utilize the reinforcement learning agent to identify the optimal design considering both resource allocation and quantization in an end-to-end fashion, for low latency and high accuracy. 
The framework can adapt to different FPGAs by changing the available resources in the cost model. 
We utilize two FPGAs XC7Z020 and XC7Z045 as the test-beds to evaluate our framework, and the explored configuration can reduce latency by 1.12$\times$-1.32$\times$ compared with the state-of-the-art Mix\&Match design.

\begin{acks}
This work was supported by Alibaba Group through Alibaba Innovative Research Program and National Natural Science Foundation of China (Grant No. 62102257).
\end{acks}

\newpage

\bibliographystyle{ACM-Reference-Format}
\bibliography{reference}

\end{document}